\PassOptionsToPackage{table}{xcolor}
\documentclass[astrosymb, twocolumn]{aastex631} %,linenumbers
\usepackage[table]{xcolor}
\usepackage{subfigure}
\usepackage{rotating}
\usepackage{mathtools}
\received{14 March 2024}
\revised{28 May 2024}
\accepted{30 May 2024}
\submitjournal{\apj}
\shorttitle{}
\shortauthors{}
\graphicspath{{./}{./}}
\begin{document}
\title{ Active Dwarf Galaxy Database I: Overlap between active galactic nuclei selected by different techniques}

\author[0000-0003-3986-9427]{Erik J. Wasleske}
\affiliation{Department of Physics and Astronomy, Washington State University, Pullman, WA 99163, USA}

\author[0000-0003-4703-7276]{Vivienne F. Baldassare}
\affiliation{Department of Physics and Astronomy, Washington State University, Pullman, WA 99163, USA}

%%%%%%%%%%%%%%%%%%%%%%%%%%%%%%%%%%%%%%%%%%%%%%%%%%%%%%%%%%%
%%
%%      Abstract
%%
%%%%%%%%%%%%%%%%%%%%%%%%%%%%%%%%%%%%%%%%%%%%%%%%%%%%%%%%%%%
\begin{abstract}

We assemble a sample of 733 dwarf galaxies ($M_{\ast} \le 10^{9.5} \text{M}_\odot$) with signatures of active galactic nuclei (AGN) and explore the intersection between different AGN selection techniques.  Objects in our database are compiled from previous studies that identify AGN in dwarf galaxies through spectroscopy, X-ray emission, infrared colors, and optical photometric variability. We apply a uniform set of AGN diagnostic tools to the database using archival data. We find that any single selection method captures no more than half of the overall AGN population, and there is a general disagreement amongst the AGN selection methods in this stellar mass regime. The largest overlap between methods is found when both methods use optical spectroscopic data. In contrast, the populations of AGN intersect the least when comparing those methods that use photometric data at different wavelengths. These results can be used to better constrain the active fraction in dwarf galaxies, which is in turn an important constraint for black hole seed formation models. In a follow-up paper, we will explore links between the effectiveness of each selection technique and host galaxy properties.

\end{abstract}

\keywords{}

%%%%%%%%%%%%%%%%%%%%%%%%%%%%%%%%%%%%%%%%%%%%%%%%%%%%%%%%%%%
%%
%%      Introduction
%%
%%%%%%%%%%%%%%%%%%%%%%%%%%%%%%%%%%%%%%%%%%%%%%%%%%%%%%%%%%%
\section{Introduction} \label{sec:intro}

At the center of virtually all massive galaxies resides a supermassive black hole (SMBH; \text{M} $\sim10^6-10^{10} \text{M}_\odot$). While the correlation between the mass of the SMBH and host galaxy properties has been well studied \citep{ferrarese2000,gebhardt2000,kormendy2013}, there remain many open questions surrounding the formation, growth, and evolution of SMBHs. There also exists a wealth of evidence for stellar mass black holes ($\sim 10 M_\odot$) created after the death of massive stars \citep{heger2003,belczynski2010}. We thus have a gap in the mass scale of black holes (BHs). The BHs that would exist in this mass range are called intermediate-mass black holes (IMBHs; ${\rm M_{BH}} \sim10^3-10^{5} \text{M}_\odot$). \cite{greene2020} provides a review of the current search for IMBHs, whose existence is predicted from formation models of SMBHs \citep{cattaneo2009,greene2020}.

There are three mechanisms considered for the formation of BH seeds in the centers of galaxies: gravitational runaways \citep{bahcall1975,begelman1978,quinlan1990}, Population III stars collapsing \citep{bond1984,madau2001}, and a direct collapse scenario \citep{haehnelt1993,loeb1994,koushiappas2004}. \cite{volonteri2010} summarizes BH formation theories and corresponding predictions for the observed population of SMBHs. They categorize the BH formation pathways as  ``light seeds"(Pop. III stars) BH formation and ``heavier seeds" (runaway and direct collapse). These models each produce a different BH seed mass within the IMBH mass range. Each mechanism results in a different BH number density over cosmic time. 

The death of Pop. III stars would result in BHs with masses $\sim 100 \text{M}_\odot$ in the very early universe ($z > 15$). The rarity of these stars \citep{madau2001} matches the number density of massive BHs today, but super-Eddington growth is required to reach some observed SMBH masses \citep{johnson2007, jiang2019}.

Direct collapse models involve the collapse of a gas cloud into a $\sim 10^{4-6}\text{M}_\odot$ solar mass black hole, potentially via an intermediate supermassive star. This requires rare conditions at high redshift to prevent gas cooling and fragmentation \citep{visbal2014}. The required values for gas density and metallicity make early proto-galaxies and enriched massive halos prime locations for this mechanism to occur \citep{bromm2003}. Models predict that this pathway will have a lower number density of BH seeds, but more massive seeds relative to the Pop. III pathway. 

Gravitational runaway models predict BH seed formation from runaway events at the center of dense stellar clusters \citep{stone2017}. This would result in BHs with mass $\sim 10^{3-4}\text{M}_\odot$. Unlike the other pathways, this process can occur at any redshift. However, \cite{breen2013} find that not all nuclear star clusters can form BHs under gravitational collapse. This pathway predicts a seed mass in between the two mechanisms described above \citep{fragione2022}.

Each of these formation pathways predicts different present-day relations between BH mass and galaxy stellar mass, BH number densities, and BH mass functions. To determine which BH formation path is dominant, it is necessary to constrain the population of BHs in the centers of galaxies across redshift and host galaxy mass. The IMBH regime is particularly constraining for distinguishing between formation scenarios.

Extrapolating from scaling relations between BH and galaxy mass, the centers of dwarf galaxies are a logical starting point to search for IMBHs. Dwarf galaxies provide a relatively pristine laboratory to study these black holes, as they do not have as extensive a merger history as more massive galaxies \citep{kormendy2013}. However, IMBHs with masses $~10^{4-5} M_\odot$ can only be detected dynamically within  $~10\text{Mpc}$. 
Studies of very nearby low-mass early-type galaxies find dynamical evidence for IMBHs in $\sim80\%$ of systems, albeit with a small sample size due to the limited number of systems that are sufficiently close \citep{nguyen2018,nguyen2020}. 

Searching for accretion signatures has become the only way to build a large, statistical sample of BHs in dwarf galaxies. Over the course of decades, a set of selection techniques have emerged for detecting central BH activity within galaxies (e.g., \citealt{ho1997, ulrich1997, berk2004, just2007, desroches2009, assef2010, stern2012}). These techniques utilize radiation from across the electromagnetic spectrum, emitted by the local environment of the accreting BH in the center of the galaxies called the active galactic nucleus (AGN). The canonical AGN consists of an X-ray corona, accretion disk, a dusty torus, and polar radio jets \citep{antonucci1993}. Divergence from this picture, especially the geometry and density of the obscuring torus, has been a point of study \citep{nenkova2008, ogawa2021}. Since these methods use optical emission lines, infrared colors, X-ray luminosities, and photometric variability across time scales and bands to identify central activity within galaxies, each separately investigates different parts of the AGN structure.  

These methods were initially established to find active SMBHs in massive galaxies. Each method has biases associated with it. Since AGN in dwarf galaxies are often lower luminosity \citep{trump2015}, their signatures are more easily obscured by the effects of star formation, gas, dust, and stellar processes within their host galaxies. Different selection techniques tend to identify different populations of dwarf galaxies as hosting AGN.

We still lack a full census of dwarf galaxies that contain AGN. Previous works have had the goal of assembling a complete population of galaxies hosting AGN using various methods. For example, \cite{ho1997} worked to compile a census of AGN in the local universe using optical spectroscopy. This method is biased against finding AGN that are heavily obscured by dust. They found that that $\approx 10\%$ of their parent population hosted AGN, with the majority of those selected residing in bulge dominated galaxies. \cite{goulding2009} sought to identify a complete sample of AGN in local galaxies using IR spectroscopy. They find that that more than half of galaxies they identify lack the signatures of an AGN in their optical spectroscopy. They state that half of all AGNs would be missing in the optical due to host galaxy dust extinction. Thus, in order to accurately capture the population of AGN in dwarf galaxies, we must first investigate the completeness and correctness of each method of AGN selection in the low-mass regime.

The pitfalls of AGN selection methods greatly affect our understanding of the fraction of actively accreting BHs in the universe. The active fraction has typically been found to be $\lesssim 1\%$ for dwarf galaxies found observationally (e.g. see \citealt{reines2013} and \citealt{baldassare2020}) where models have found the active fraction to potentially be as high as $5-22\%$ \citep{pacucci2021}. A well constrained value of the active fraction will give a firm lower bound on the occupation of BHs in the center of dwarf galaxies. Alongside the scaling relations for BH-galaxy mass, a true value of the active fraction will provide constraints for the seeding models mentioned above. Thus the active fraction of BHs observed today is a important value in the work towards our understanding of the formation and evolution of BHs and galaxies throughout cosmic time.

Cosmological zoom-in simulations incorporate different initial BH seed masses, criteria for seeding halos, and BH accretion and feedback prescriptions (ie. see \citealt{habouzit2021} and the works of \citealt{genel2014,nelson2017,dubois2014,schaye2014,dave2019}). Each simulation uses different sub-grid physics to model the physical processes within these galaxies. The observational constraints of these simulations often come from single-method papers (e.g., X-ray active fraction in \citealt{birchall2020}). \cite{haidar2022} expands on the work of \cite{habouzit2021} by studying the agreement of the BH populations in each simulation with the observed population of BHs in low mass galaxies. Differences in prescriptions for BH formation, evolution, and feedback affect the derived BH occupation and active fractions.

Results from our work can provide revisions to single-method active fraction estimates for better comparison to simulations such as these. A multi-wavelength active fraction provides a much better picture of the overall population to compare to simulations and constrain the occupation fraction. Better comparisons will yield insights into the preferred sub-grid physics needed to replicate the observational results.

Here, we present a compilation of known AGN in dwarf galaxies selected by different techniques. We then uniformly apply each method of identifying AGN to the entire sample. We compare results from each method and study the overlap of different techniques. In a follow-up paper, we will explore whether host galaxy properties (e.g., star formation rate, metallicity, morphology) impact the selection of an AGN with a given technique. 

This paper is organized as follows: Section \ref{sec:data} discusses the source populations used to construct our database of known AGN in dwarf galaxies; Section \ref{sec: Consistent Application of AGN selection tools} describes the uniform application of AGN selection tools to our database; Section \ref{sec:results} examines the results of these selection tools and the overlap between different techniques. Section \ref{sec:discussion} will consider these results, with future work investigating the biases and physical reasoning of the results. We assume a $\Lambda$CDM cosmology with parameters $h=0.7$, $\Omega_m=0.3$ and $\Omega_\lambda=0.7$ \citep{spergel2007}.

%%%%%%%%%%%%%%%%%%%%%%%%%%%%%%%%%%%%%%%%%%%%%%%%%%%%%%%%%%%
%%
%%      Data
%%
%%%%%%%%%%%%%%%%%%%%%%%%%%%%%%%%%%%%%%%%%%%%%%%%%%%%%%%%%%%
\section{Data} \label{sec:data}

We first construct a database of known dwarf galaxies with AGN signatures across the electromagnetic spectrum. We describe each AGN selection method and corresponding papers below. The individual samples of dwarf galaxies were cross matched using a 2-D sky separation to account for any overlapping objects. For each object we track the original source papers and selection methods used to identify it as a dwarf AGN. 

We compile the host galaxy stellar mass, redshift, and optical g and r band magnitude values for each object. For objects whose source paper catalogs lacked any of these values, the catalog positions were queried in SIMBAD \citep{wenger2000} to fill in these missing values. For our database, we impose a host galaxy stellar mass cut of $\text{log}\left(\text{M}_* \right) \leq 9.5$ to the resulting populations to match \cite{reines2013} and \cite{sartori2015}. This mass cut is lower than used in some of our source papers, but is chosen since changes in the occupation fraction for different formation scenarios of BHs are most evident within the range $8.5 < \text{log}\left(\text{M}_* \right) < 9.5$ (see \citealt{greene2020} for a review on formation pathways). 

The initial database contains 733 galaxies, with redshifts $0.001 \leq \text{z} \leq 1.5$, host stellar masses down to $\text{log}\left(\text{M}_* \right) = 6.13$. Optical g and r band magnitudes range from $12.1 < m_\text{g} < 27.4$ and $11.6 < m_\text{r} < 23.2$ respectively. A summary of the population of galaxies taken from each study is given in Table \ref{tab: sources table }.

\subsection{Optical Spectroscopic Selected Candidates} \label{subsec: Data - spec}
%- Reines+13 : BPT
\cite{reines2013} searched for optical spectroscopic AGN signatures in ~25,000 nearby ($z<0.055$) dwarf galaxies in the NASA-Sloan Atlas, a catalog of galaxies with Sloan Digital Sky Survey (SDSS) spectroscopy. They build a pipeline for fitting the narrow and broad emission lines of dwarf galaxies. Using their measured fluxes, they identified 136 active dwarf galaxies using the BPT optical spectroscopic diagnostic diagram \citep{baldwin1981} as their primary method. The 136 active dwarf galaxies include both BPT AGN and composite galaxies. 10 of these 136 also had broad H$\alpha$ emission.

%- Moran+14 : BPT
\cite{moran2014} identified 28 AGN in nearby ($\leq80 \text{Mpc}$) dwarf galaxies from SDSS, with Seyfert-type emission. They identified a low-luminosity population  with the majority of galaxies (all but 2) being narrow-line objects. After our mass cut, we keep 16 of these objects as part of our database, four of which are unique to \cite{moran2014}.

%- Sartori+15 : BPT + HeII
\cite{sartori2015} uses multiple methods to search for active galaxy candidates. Starting from a parent population of $~48,000$ low-mass $(M_* \leq 10^{9.5} M_\odot)$ local $(z< 0.1)$ galaxies from SDSS DR7 \citep{abazajian2009}, they identify 336 AGN candidates. They implement two spectroscopic selections: the standard BPT diagram and the [\ion{He}{2}] emission diagram \citep{shirazi2012}. This identified 48 galaxies in the BPT diagram's AGN-Seyfert region and 121 galaxies in the AGN region of the [\ion{He}{2}] diagram. We include 45 of the BPT and 117 of the [\ion{He}{2}] selected in our database, noting 42 of these objects are identified by other surveys or other techniques in \cite{sartori2015}. The remaining AGN candidates in their population were identified by additional techniques described below.

%- Chiligarian+ : broad-line
\cite{chilingarian2018} carried out an analysis of archival and follow-up spectra of galaxies found through the data mining of wide-field sky surveys to search for signatures of type 1 broad-line AGN. They found a sample of 305 galaxies. They suggest these galaxies host black holes with masses from $3 \times 10^4 \text{M}_\odot$ to  $2 \times 10^5 \text{M}_\odot$. We include 62 of these objects after the mass cut, six of which are identified by other surveys.

%- Salehirad+22 : BPT, HeII, coronal
\cite{salehirad2022} identifies a population of galaxies with spectroscopic signatures of AGN within the Galaxy and Mass Assemble (GAMA) survey DR4 \citep{liske2015, driver2022}. They select their candidates using four techniques: BPT and [\ion{He}{2}] diagnostic diagrams and selection via the coronal lines of [\ion{Fe}{10}] and [\ion{Ne}{5}]. 73 of the 338 galaxies in their sample meet our stricter mass cut and are included in our sample. 

\subsection{X-ray Selected Candidates} \label{subsec: Data - x-ray}
%- Birchall+20 : 
\cite{birchall2020} selected AGN in low-redshift ($z \leq 0.25$) dwarf galaxies via their X-ray emission. They started with a population of 4331 dwarf galaxies within the SDSS DR8 and 3XMM DR fields. They found 61 of these dwarf galaxies to to have a bright X-ray counterpart to their nucleus through a rigorous position matching/separation criteria and ensuring point-like sources by limiting the extent of the X-ray source to $<10 \: \text{arcsec}$. These objects additionally met a criteria of their observed X-ray luminosity being $3\times$ the expected X-ray luminosity. The expected X-ray luminosity is defined as the sum of the likely X-ray emission from X-ray Binaries (XRB) and hot interstellar gas. After our mass cut, we include 33 of these galaxies in our database, with seven of these selected by additional surveys.

\subsection{Infrared Selected Candidates} \label{subsec: Data - IR}
%- Sartori+15 : 
Using Wide-field Infrared Survey Explorer (WISE; \citealt{wright2010}) magnitudes, \cite{sartori2015} implemented the color-color selections of \cite{jarrett2011} and \cite{stern2012}. 189 of their galaxies made at least one of these cuts. Of these 189, only 3 galaxies are also selected by their BPT diagram and 4 by the [\ion{He}{2}] diagram. In comparing the population selected by each of their techniques, \cite{sartori2015} notes that these 189 mid-IR candidates are bluer and less massive than their spectroscopic selected candidates, and all lack broad H$\alpha$ lines. We include all 189 from this IR selection in our database, while six of these objects are identified in the other surveys. 

\subsection{Variability Selected Candidates} \label{subsec: Data - variability}
%- Baldassare+18 :
\cite{baldassare2018} constructed SDSS g-band light curves via difference imaging of $\sim28,0000$ low-redshift galaxies to search for nuclear variability. They compared these light curves to a damped random walk model to identify which galaxies have significant AGN-like variability. This method selected 135 galaxies with AGN-like nuclear variability. Five of these objects are included in our database after our mass cut. They found that the low-mass variable galaxies tended to fall in the star forming region of the BPT diagram.

%- Baldassare+20 :
Using Palomar Transient Factory (PFT) R-band observations, \cite{baldassare2020} presented an analysis of $\sim50,000$ galaxies  with $\text{z} < 0.055$ to search for nuclear variability. They use difference imaging aperture photometry to construct light curves and identify galaxies whose nuclear variability is indicative of AGN via comparison to a damped random walk model, similar to that of \cite{baldassare2018}. They find 237 galaxies with AGN-like variability and stellar masses less than $10^{10} \text{M}_\odot$. After our mass cut, 105 of these galaxies are included in our database, with three being cross-listed with other surveys.

%- Burke+22 : 
\cite{burke2022} searched for AGN via optical variability in three deep fields of the Dark Energy Survey (DES). They also use difference imaging aperture photometry to construct light curves, finding 706 AGN within $z<1.5$ and $4.64 \text{deg}^2$ of the field. Using spectral energy distribution modeling (SED), the made stellar mass estimates of these objects. This analysis found 34 objects to have a reliable mass estimate with a value $<10^{9.5} \text{M}_\odot$, all of which we included in our database.

%- Wasleske+22 :
\cite{wasleske2022} searched for AGN candidates via ultraviolet variability. They constructed near-ultraviolet light curves from the Galaxy Evolution Explorer (GALEX) Time-Domain Survey \citep{gezari2013} for ~2,000 objects from the NASA Sloan Atlas catalog. Of the 48 high-probability candidates, six were also identified as low-mass ($<10^{10} \text{M}_\odot$), which we include 4 of them in our database after our stricter mass cut.

\begin{table*}[t]
    \centering
    \textbf{Sources of Objects\\}
    \begin{tabular}{c | c c c c}
\hline
Source & Selection Method(s) &  Number of Included Objects &  Overlapping Objects & Unique Objects\\
%\hline
%    &   \%   & \% & \%  \\
\hline
\hline

\cite{reines2013}&  Spectral   & 136  & 47 & 89 \\
\cite{moran2014}&  Spectral   & 16  &  12 & 4\\
\cite{sartori2015} & Spectral \& IR & 331  & 42 & 289\\
\cite{chilingarian2018} & Spectral & 62 & 6 & 56\\
\cite{salehirad2022} & Spectral & 73 & 0 & 73\\
\cite{birchall2020} & X-ray & 33 & 7 & 26\\
\cite{baldassare2018} & Variability & 5 & 0 & 5\\
\cite{baldassare2020} & Variability & 105 & 3 & 102\\
\cite{burke2022} & Variability & 34 & 0 & 34\\
\cite{wasleske2022} & Variability & 4 & 0  & 4\\

\hline

\end{tabular}
    \caption{Summary of the compiled AGN candidates from the source papers outlined in Section \ref{sec:data}. Columns identify the type of selection method used, the number of galaxies found in the study, and how many of these objects overlap with the other studies. The last column gives the number of galaxies identified only by that study within this set of works.}
    \label{tab: sources table }
\end{table*}

%%%%%%%%%%%%%%%%%%%%%%%%%%%%%%%%%%%%%%%%%%%%%%%%%%%%%%%%%%%
%%
%%      Analysis
%%
%%%%%%%%%%%%%%%%%%%%%%%%%%%%%%%%%%%%%%%%%%%%%%%%%%%%%%%%%%%
\section{Consistent Application of AGN selection tools} 
\label{sec: Consistent Application of AGN selection tools}

In this section we describe the uniform application of a set of AGN indicators to the 733 objects in our database. Since the parent populations used for each of the studies above are different, not all of the galaxies in our database were originally analyzed for the same AGN signatures.  We generate our own classifications for the 733 galaxies in our initial database, based on the outcome of each AGN selection method. Below we discuss the application of each of these methods to all galaxies within the database to create these sub-samples. We compare the results from each method in Section \ref{sec:results}.

%%%
\subsection{Optical Spectroscopic Diagnostics} \label{subsec: Analysis - emission line}
We collect all available spectra from SDSS and GAMA for our sample. We queried SDSS via \texttt{astroquery}'s \texttt{query\_region} function using the object's position from the source paper. For the GAMA survey, we queried DR4 for GAMA objects taken from \cite{salehirad2022} \footnote{http://www.gama-survey.org/}.

We found 674 objects in our database that have a spectrum in either SDSS or GAMA. We found no spectrum for one object originally selected via BPT (NSA 88154), two selected via [\ion{He}{2}] (Sartori 11212 and Sartori 36713), four via broad lines, one selected one selected via mid-IR (Sartori 306717) and 51 identified via variability. These variability identified targets, from \cite{burke2022} are from the southern fields of the DES. Neither SDSS nor GAMA has coverage within these fields.

We model the optical spectra using \texttt{PyQSOFit} software \citep{guo2018}. We model the emission lines of H$\alpha$$\lambda 6563$, [\ion{N}{2}]$\lambda 6549$, [\ion{N}{2}]$\lambda 6585$, [\ion{S}{2}]$\lambda 6718$, [\ion{S}{2}]$\lambda6732$, H$\beta$$\lambda 4861$, [\ion{O}{3}]$\lambda 4959$ , [\ion{O}{3}]$\lambda 5007$, [\ion{He}{2}]$\lambda 4687$, [\ion{O}{1}]$\lambda 6300$,[\ion{O}{1}]$\lambda 6363$, and [\ion{Fe}{10}]$\lambda 6374$. For each spectrum, we perform two separate fits. In the first fit, we include only narrow line emission. In the second fit, we include broad components for H$\alpha$ and H$\beta$. Each broad component is modeled as the sum of two Gaussians, each whose width must be $\geq 500 \text{km/s}$. We select either the narrow line only fit or the narrow + broad fit based on the reduced $\chi ^2$ of the H$\alpha$ complex. If the inclusion of the broad component decreases the reduced $\chi ^2$ by 20\% or more, the narrow + broad fit is selected. Otherwise, we use the narrow line only fit model. This gives us an intial sample of galaxies with broad H$\alpha$. We elaborate on this sample in Section 3.1.4.

Our analysis based on the measured line fluxes is described below. 

\subsubsection{BPT} \label{subsubsec: Analysis - BPT}
\begin{figure}
  \centering
    \includegraphics[width=0.50\textwidth]{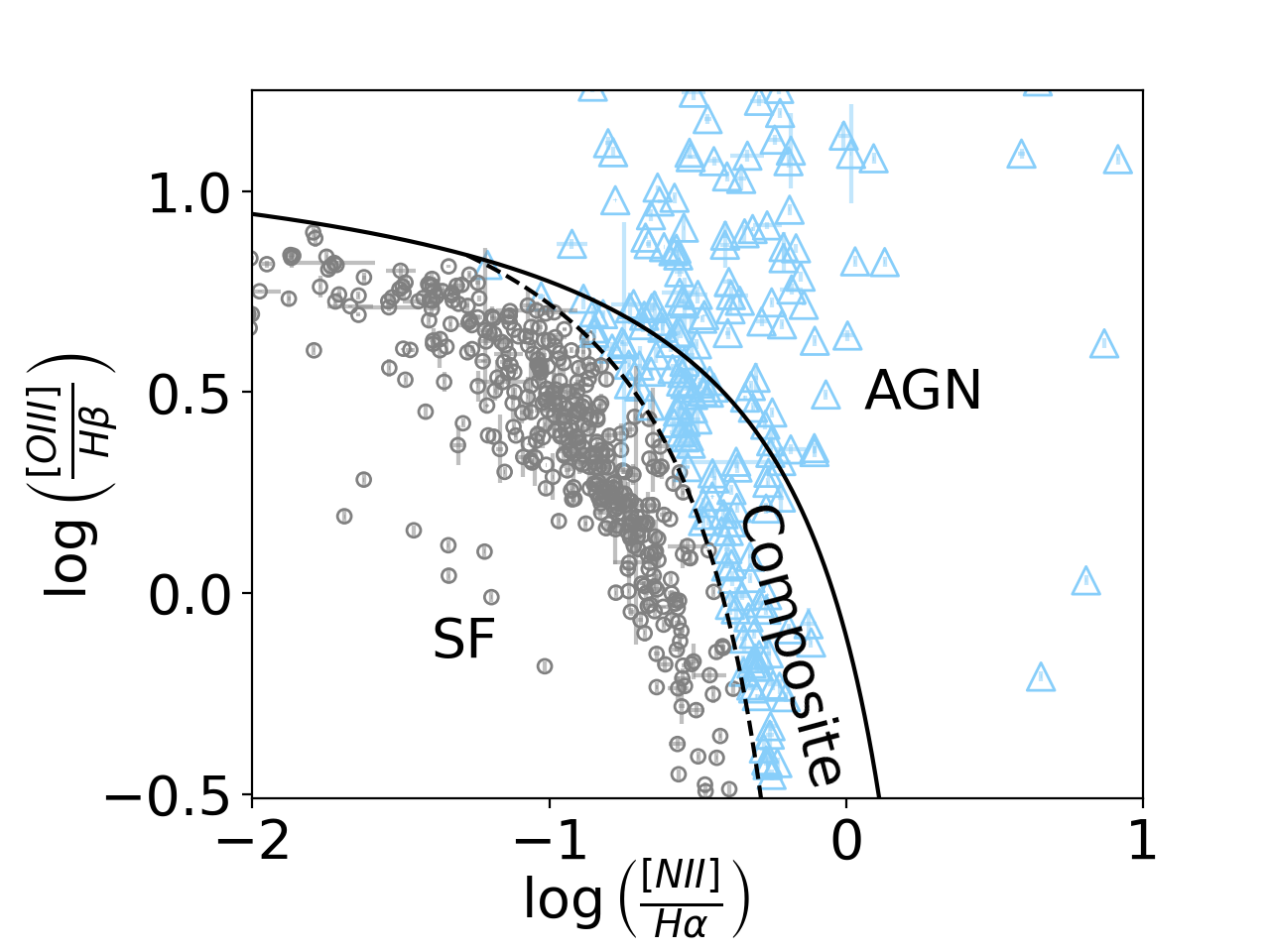}
  \caption{The BPT diagram for the 674 galaxies in our database with optical spectroscopy. Classification lines separating these regimes are from \cite{kewley2001, kewley2006} and \cite{kauffmann2003}. The black solid line from \cite{kewley2006} separates objects dominated by AGN from those with some star formation (SF). The black dashed line in the Figure, taken from \cite{kauffmann2003}, distinguishes between those objects with composite optical signatures and those dominated solely by SF. Light blue triangles are those selected as having active BH signatures. Grey circles are those objects whose emission is dominated by SF instead of BH activity.}
  
  \label{fig: BPT diagram}
\end{figure}

We first investigate location of the 674 galaxies with available spectra on the BPT diagram (Figure \ref{fig: BPT diagram}). The most commonly used emission line diagram is the [\ion{N}{2}]/H$\alpha$ vs. [\ion{O}{3}]/H$\beta$ define in \cite{baldwin1981}, which we refer to as the BPT diagram. This diagram uses narrow emission line ratios to separate galaxies containing actively accreting BHs from those dominated by star formation (SF). The SF regime is dominated by emission from gas photoionized by O and B type stars. The AGN regime is a result of harder continuum emission from AGN accretion disk photons ionizing the interstellar medium of the galaxy. The result is a separation into two distinct regions in the [\ion{N}{2}]/H$\alpha$ vs. [\ion{O}{3}]/H$\beta$ space. In between the theoretical maximum starburst line and empirical AGN line is the composite region, where galaxies are expected to have contributions from both SF and an AGN. Galaxy metallicity also affects where objects fall on the BPT diagram, as lower metallicity causes a higher [\ion{O}{3}]/H$\beta$ and lower [\ion{N}{2}]/H$\alpha$ value \citep{groves2006,reines2013}. 

Using the BPT diagram, we classify objects whose narrow emission lines place them above the theoretical maximum SF line as ``BPT Selected" since they should have at least some contribution from an active BH. Of the 674 galaxies with fitted spectra, 268 are classified as ``BPT-selected AGN". They are shown as the light blue triangles in Figure \ref{fig: BPT diagram}.

\subsubsection{[\ion{O}{1}] and [\ion{S}{2}] } \label{subsubsec: Analysis - OI and SII}

We also analyze the sample using the [\ion{O}{1}] and [\ion{S}{2}] diagrams. The [\ion{O}{1}] and [\ion{S}{2}] diagrams are similiar to the [\ion{N}{2}]/H$\alpha$ vs. [\ion{O}{3}]/H$\beta$ diagram, and are also defined in \cite{baldwin1981}. These, often found in unison with the [\ion{N}{2}] BPT diagram, are used to separate AGNs into either Seyfert or low-ionization emission line ratio (LINER) type galaxies. LINERs have spectra that contain weak ionization lines with varying galaxy morphologies. There are open questions as to whether ionization in LINERs is from a BH or stellar processes \citep{kewley2006, agostino2021}. 

\begin{figure}[h]
    \centering
    \subfigure{\includegraphics[width=0.50\textwidth]{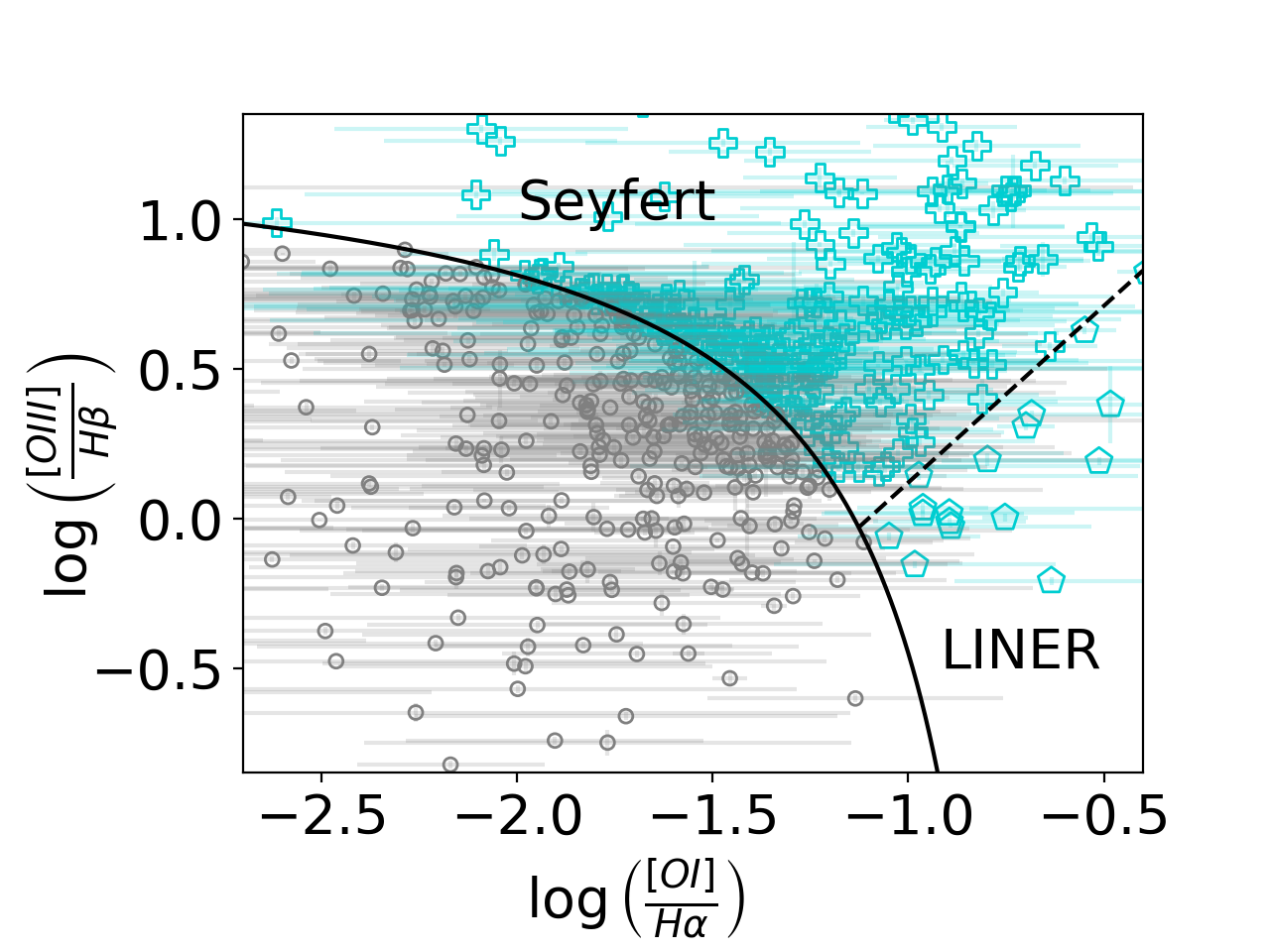}}%\quad
    \hspace{1em}% Space between image A and B
    \subfigure{\includegraphics[width=0.50\textwidth]{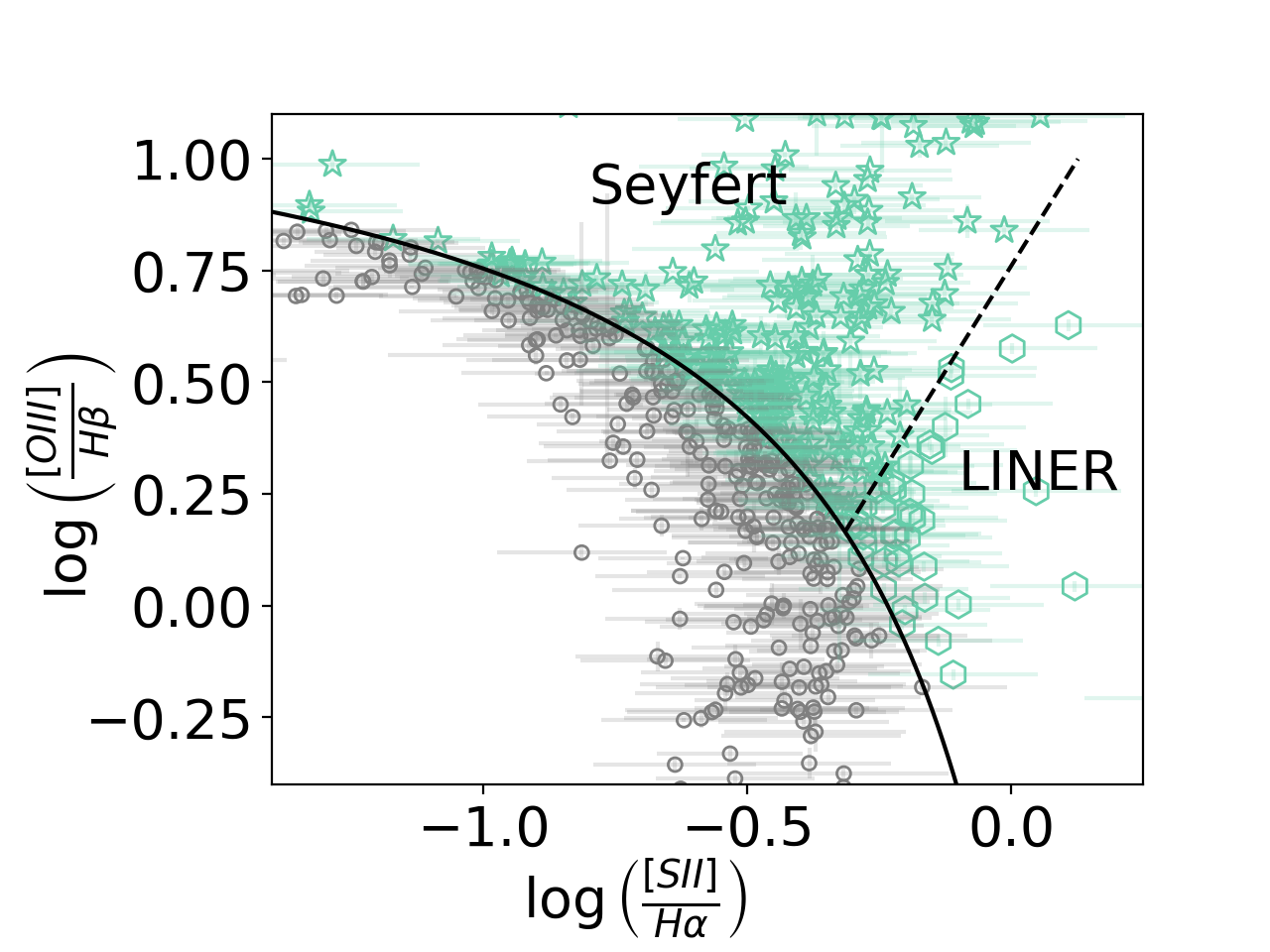}}
    \caption{ Optical diagnostic diagrams similar to the BPT diagram, but with  [\ion{O}{1}] (\emph{top}) and [\ion{S}{2}](\emph{bottom}) instead of [\ion{N}{2}] . The solid black line separates SF objects from Seyfert and LINERs. The black dashed line here separates Seyferts from LINER-type galaxies. Grey points represent galaxies not selected as AGN by this diagram. Teal plus signs mark [\ion{O}{1}]-Seyferts and teal pentagons mark th  [\ion{O}{1}]-LINERs in the top diagram. [\ion{S}{2}] Seyferts and [\ion{S}{2}] LINERs are marked by seafoam green stars and hexagons, respectively, in the bottom plot. 
    }
    \label{fig: OI SII diagrams}
\end{figure}

The [\ion{O}{1}] diagram uses the ratios [\ion{O}{1}]/H$\alpha$ vs. [\ion{O}{3}]/H$\beta$ as [\ion{O}{1}]/H$\alpha$ is more sensitive to the radiation field's hardness \citep{kewley2006}. We show this diagram in the top panel of Figure \ref{fig: OI SII diagrams}. We find that 270 of the galaxies in our database with fitted optical spectra are classified as Seyferts and 36 as LINERs on this [\ion{O}{1}] diagram. We use these classifications as additional methods of AGN selection.

Analogously, the [\ion{S}{2}] diagram uses the narrow line emission ratios of [\ion{S}{2}]/H$\alpha$ vs. [\ion{O}{3}]/H$\beta$. This diagram also separates our AGN into the same classes as the [\ion{O}{1}] diagram whilst using the stronger [\ion{S}{2}] emission line. The bottom panel of Figure \ref{fig: OI SII diagrams} shows our sample on this diagram. We find 290 galaxies classified as Seyferts and 61 as LINERs through this diagnostic, which we add to our sub-samples of AGN selection methods for further analysis below.

\cite{reines2013} found that their AGN from the standard BPT diagram are majority Seyferts on these two diagrams, with the Composites from the BPT diagram falling all across these two diagrams. Thus the diversity of classifications that we find from these three methods are consistent with their results for a smaller sample. We discuss the intersection of these methods further in Section \ref{sec:results}.

%%%

\subsubsection{[\ion{He}{2}] diagram} \label{subsubsec: Analysis - HeII}
We also study diagnostics using the [\ion{He}{2}] emission line. Part of our PyQSO fitting was to measure the emission from [\ion{He}{2}] 4686\AA. 

We use the established [\ion{He}{2}] selection criteria from \cite{shirazi2012}. Ionized helium has a high ionization potential ($54.4 \text{eV}$), thus only phenomena that produce hard radiation, such as AGN and young stellar populations, would produce this emission \citep{sartori2015}. As [\ion{He}{2}] is a line produced in the near-neutral interstellar medium, it can help identify weaker AGN with low metallicities. 

\begin{figure}[h]
  \centering
    \includegraphics[width=0.5\textwidth]{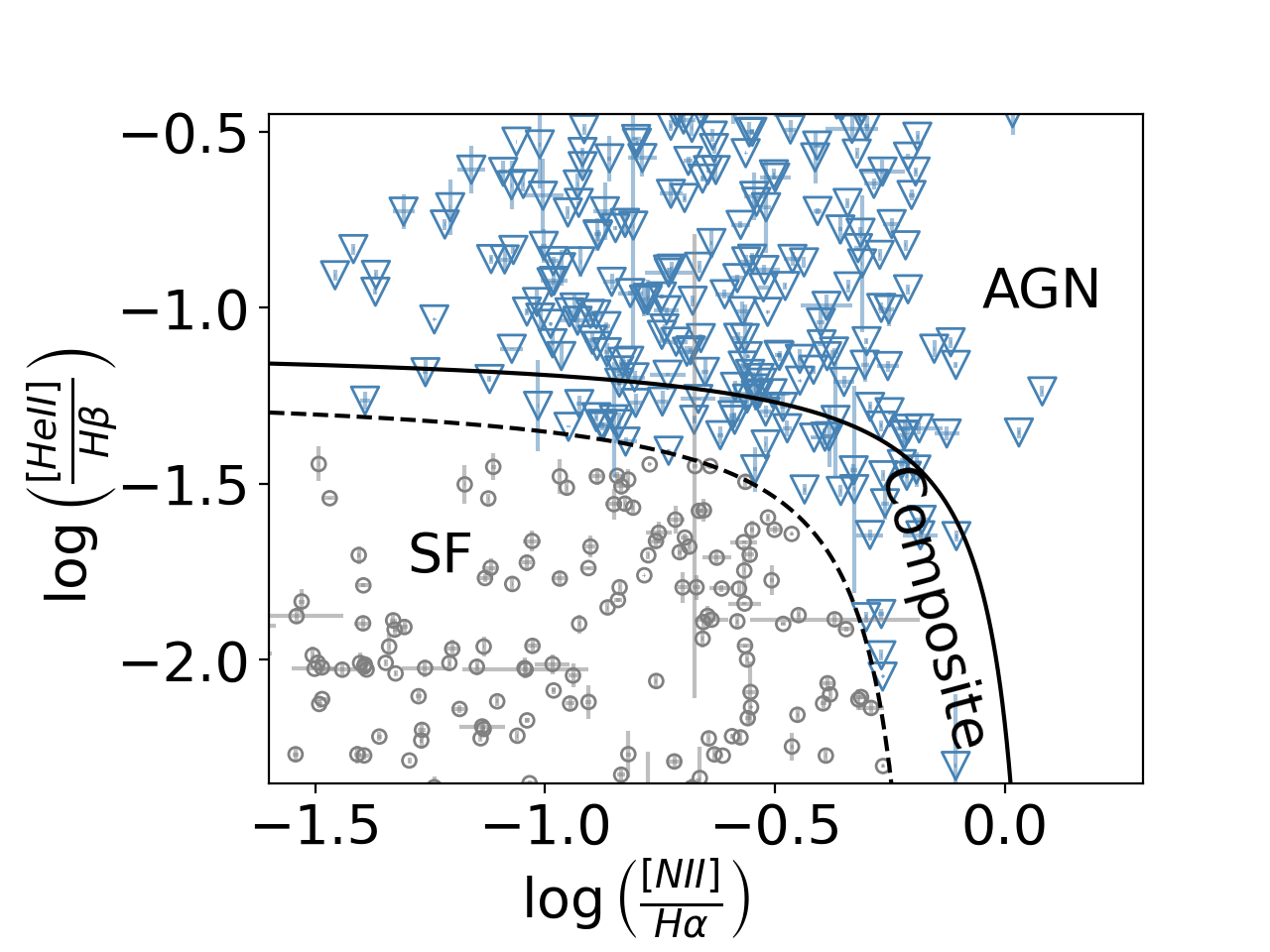}
  \caption{The [\ion{He}{2}] optical diagnostic applied to galaxies with fitted spectra. The solid black line represents the theoretical maximum value for starburst galaxies. \cite{shirazi2012} established this by adopting the stellar population models of \cite{charlot2001} and setting a model grid with inputs for metallicity, redshift, ionization parameter, dust attenuation and the dust-to-metal ratio. The dashed line is the empirical separation between SF galaxies and those that are composite and AGN. }
  \label{fig: HeII diagram}
\end{figure}

Figure \ref{fig: HeII diagram} shows our sample on the [\ion{He}{2}] diagram. We select those in the Composite and AGN regimes as our ``[\ion{He}{2}] Selected" candidates. 361 objects out of the 674 galaxies with fitted spectra are classified as ``[\ion{He}{2}] selected" AGN, 93 more galaxies than identified by the standard BPT diagnostic. 

\subsubsection{Broad-line Emission} \label{subsubsec: Analysis - broad-line}
 As mentioned above, we run \texttt{PyQSOFit} twice to select either a narrow emission line model or a narrow+broad line model. For the narrow+broad line model, we included a broad component for the H$\alpha$ and H$\beta$ lines using two components in addition to the narrow emission lines.

We found 115 galaxy spectra met the requirement of the H$\alpha$ complex $\chi ^2$ being reduced by 20\% or with the addition of broad components. We then filter for the quality of these fits by making cuts to the signal-to-noise ratio (SNR) and Full Width Half Maximum (FWHM) of the broad H$\alpha$ line. We define the SNR for the broad H$\alpha$ line as the measured flux over the error of this flux measurement for the total broad line. If the SNR was greater or equal to 3 and the FWHM was $500 \text{km/s}$ or greater for the broad H$\alpha$, then the object was kept for continuing analysis of its broad lines. After these quality cuts, we were left with 73 broad-line candidates. 

\begin{figure*}[t]
  \centering
    \includegraphics[width=\textwidth]{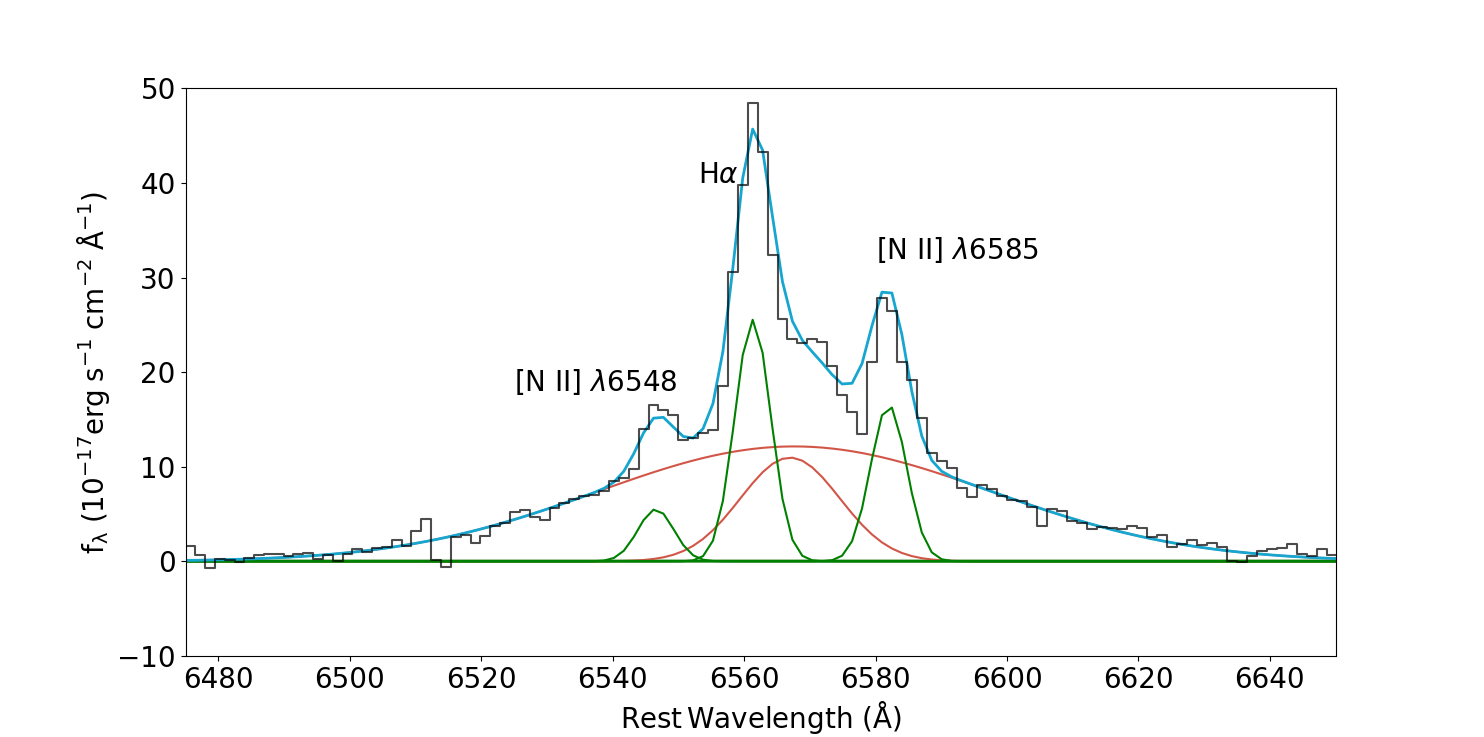}
  \caption{The broad+narrow line fitting of H$\alpha$-[\ion{N}{2}] complex of the galaxy J022253.62-042929.22. This galaxy was originally found via optical variability in the work of \cite{burke2022}. Continuum subtracted spectra is given in black, with the total model as the blue line. The three green lines are single Gaussian fits, corresponding to the narrow-line emission of [\ion{N}{2}]$\lambda$6548, H$\alpha$, [\ion{N}{2}]$\lambda$6584. The red lines give each of the two components of the Gaussian fit to the broad H$\alpha$ emission.}
  \label{fig: broad-line example}
\end{figure*}

We inspected the broad-line fit of these 73 objects by eye as our final criteria. Our inspection looked for closeness of fit and over-fitted broad lines (leading to very low reduced $\chi ^2$ for the H$\alpha$ complex).  We found that 25 fits appeared reliable.  An example of a broad H$\alpha$ fit is shown in Figure \ref{fig: broad-line example}. We searched for evidence of ionized gas outflows in these 25 broad-line AGN by modeling asymmetric wings of the [\ion{O}{3}] emission \citep{singha2022}. Of the 25 objects, 3 appear to have redshifted or blueshifted [\ion{O}{3}] wings. J022253.62-042929.22 has a redshifted wing whose width is $162 \text{km/s}$. The other two objects, NSA 15235 and NSA 104527, have blue-shifted wings in their [\ion{O}{3}] emission. The amplitude ratio of the wing component to the narrow component is $<1/5$ for both galaxies. With these components being relatively narrow or low amplitude, we don't expect this outflow signature to impact the broad H$\alpha$ fit.

With the remaining 25 bona fide broad-line AGN candidates, we calculated the central BH mass using the broad H$\alpha$ emission. We use the $M_{BH}$ equation from \cite{reines2013}, given as 
\begin{equation}
    \begin{aligned}
    \text{log} \left( \frac{\text{M}_{BH}}{\text{M}_\odot} \right) = \: & 0.47 \: \text{log} \left( \frac{\text{L}_{H\alpha}}{10^{42} \text{erg s}^{-1}} \right) \\
    & + 2.06 \: \text{log} \left( \frac{\text{FWHM}_{H\alpha}}{10^{3}\text{km s}^{-1}}\right) + 6.57  
    \end{aligned}
    \label{equation: M_BH}
\end{equation}
where $\text{L}_{H\alpha}$ is the luminosity of the  broad line emission and $\text{FWHM}_{H\alpha}$ is the full-width half-max of the broad component. This measurement has a systematic uncertainty of 0.3 dex. We find a range of $ 4.6 \leq \left( \frac{\text{M}_{BH}}{\text{M}_\odot} \right) \leq 7.0$ from this equation for our remaining 25 broad-line candidates. We discuss scaling relations of these values in Section \ref{subsubsec: Results - broad-line}.

%%%
\subsubsection{Coronal emission lines} \label{subsubsec: Analysis - Exclude Coronal}

Within our database, 29 objects were initially identified by \cite{salehirad2022} via their [\ion{Fe}{10}] coronal line emission. They measure $\text{L}_{[\text{Fe X}]}$ luminosities of $\approx 10^{38-40} \text{erg s}^{-1}$ for this sample. These luminosities are higher than expected from supernovae \citep{molina2021} and indicative of a TDE or AGN. 

The [\ion{Fe}{10}]$\lambda 6374$  is usually hard to measure as it is very weak and can sometimes be blended with the adjacent [\ion{O}{1}]$\lambda 6363$.  To manage this issue, \cite{molina2021} first required a SNR of [\ion{O}{1}]$\lambda 6300$ to be $\geq 3$. They then used this line as a template for [\ion{O}{1}]$\lambda 6363$, constraining the widths to the same velocity and enforcing that the ratio of  [\ion{O}{1}]$\lambda 6300$/[\ion{O}{1}]$\lambda 6363$=3. 

Contamination of neighboring lines to [\ion{Fe}{10}] can also lead to misinterpretation of this emission, as in the case found by \cite{herenz2023}. They found that the supposed [\ion{Fe}{10}] of SDSS J094401.87-003832.1 \citep{reefe2023} is actually the [\ion{Si}{2}]$\lambda 6371$ line. They followed the methodology of \cite{molina2021} to model the [\ion{O}{1}]$\lambda 6363$. They additionally modeled the [\ion{Si}{2}]$\lambda 6313$. After removing these lines, the residual spectrum found line consistent with [\ion{Si}{2}]$\lambda 63147$ and [\ion{Si}{2}]$\lambda 6371$. After creating a template from \ion{Si}{2}]$\lambda 63147$ to fit the [\ion{Si}{2}]$\lambda 6371$ line, the removal of the [\ion{Si}{2}]$\lambda 6371$ emission left no remaining trace of a possible [\ion{Fe}{10}] signal. \cite{herenz2023} thus cautions the use of [\ion{Fe}{10}] emission as a tracer for AGN, as a [\ion{Si}{2}]$\lambda 63147$ in the spectrum leads to [\ion{Si}{2}]$\lambda 6371$ emission that can pollute the [\ion{Fe}{10}] signal.

We checked for coronal emission by fitting [\ion{Fe}{10}] narrow line and checking its SNR value. When we applied a SNR cut of  $\geq 3$ to the 663 objects with [\ion{Fe}{10}] fitted emissions, no objects met the criteria. We apply this strict criteria as a quality-of-fit check, due to the issues mentioned above. We note that the PyQSOFit routine did not use the methodology applied by \cite{molina2021}. Due to the low SNR of the measured [\ion{Fe}{10}] lines in our fits, we do not pursue this selection technique further for our sample.

The [\ion{Ne}{5}]$\lambda 3427$ is also considered to be an effective tool for identifying AGN that are missed by other diagnostics \citep{vergani2018}. 

We checked for [\ion{Ne}{5}] emitters within our database. Given that the minimum wavelength of GAMA spectra is 3750\AA, a minimum redshift of $\text{z} = 0.094$ is required to have this [\ion{Ne}{5}] line in their spectra.  We found that of those objects with GAMA spectra, 25 meet this redshift cut. These objects were all originally found in \cite{salehirad2022}. In addition to these 25 GAMA objects, 30 objects were at a high enough redshift to search for this line within SDSS spectra. These galaxies are at $\text{z} > 0.108$, given the minimum wavelength of 4000\AA \: for SDSS spectra. All 30 were originally identified via variability, with 28 of them from \cite{burke2022} and one from \cite{baldassare2018} and \cite{baldassare2020} each. 27 of the 30 objects did not have any available SDSS spectra. We note that the blue end of the spectra is often too noisy for proper fitting, thus meeting this redshift cut alone does not indicate that this line can be fitted.

From the 28 objects with spectra and sufficiently high redshifts, only one object had a [\ion{Ne}{5}] emission that could be fitted. This object, J022253.62-042929.22, has a [\ion{Ne}{5}] line with a luminosity of $4.38 \pm 2.26 \times 10^{40}$ erg/s. Even with this bright detection, the SNR is too low ($<2$) to make quality cuts for a tracer of AGN activity.

%%%

\subsection{WISE IR} \label{subsec: Analysis - WISE IR}

We queried the WISE \citep{wright2010} All-Sky Source Catalog for IR photometry in the four WISE bands, with wavelengths of 3.4, 4.6, 12 and 22 $\mu \text{m}$ for W1, W2, W3, and W4 respectively. This catalog requires a $\text{SNR} > 5$ and contains photometry for over 500 million point-like and resolved sources.

Only 80 of our 733 galaxies did not have available WISE photometry for these bands. Source matching was done using the closest matching source to the galaxy position from the literature within a $2\text{"}$ 2-D sky distance. 

\begin{figure}[h]
  \centering
    \includegraphics[width=0.50\textwidth]{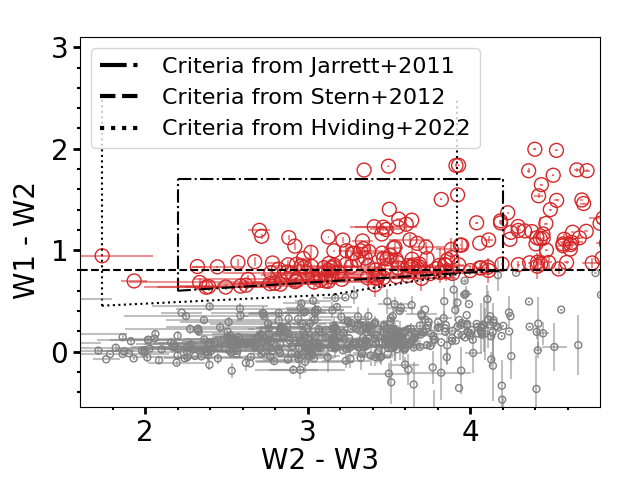}
  \caption{The WISE color-color diagram for all available WISE data for galaxies in our database. The selection criteria of \cite{jarrett2011}, \cite{stern2012}, and \cite{hviding2022} are given as black dot-dash, dashed, and dotted line respectively. Points meeting any of these criteria were selected as IR-AGN colored in red. Those not selected are shown in grey.
  }
  \label{fig: WISE diagram}
\end{figure}

We use WISE color-color diagrams as the IR AGN selection method. We use the selection criteria of \cite{jarrett2011}, \cite{stern2012}, and \cite{hviding2022}. In Figure \ref{fig: WISE diagram}, we show the WISE color-color plot for all objects in our database with WISE photometry. The criteria from \cite{stern2012} identifies objects as AGN if they have a value of $\text{W1} - \text{W2} > 0.8$. The criteria from \cite{jarrett2011} and \cite{hviding2022} define more complex regions where objects that fall within it on the color-color diagram are selected as IR AGN.

The \cite{hviding2022} selection criteria were created to identify objects at lower luminosities and higher obscuration than traditional selection methods such as \cite{jarrett2011} and \cite{stern2012}. These traditional methods are designed to select AGN that dominate the host galaxy emission in the mid-IR, as they are sensitive to detecting power law emission from hot dust components. However, \cite{hainline2016} demonstrates the possibility of star-forming galaxies diluting the WISE color-selected AGN.

Nevertheless, given the popularity of this AGN diagnostic, we include it in this study in order to explore the overlap between IR-color selected objects and those identified through other techniques. The WISE color-color diagram for our sample is shown in Figure \ref{fig: WISE diagram}. 201 out of 653 objects are selected as WISE AGN.  

%%%

\subsection{X-ray Emission} \label{subsec: Analysis - X-ray}
We measure X-ray emission from archival data from the Chandra X-ray Observatory and XMM-Newton Source Survey \citep{webb2020}. 

\subsubsection{Chandra X-ray Observatory} \label{subsubsec: Analysis - CXO}
To calculate the X-ray emission of objects with the Chandra X-ray Observatory data, we followed the analysis outlined in \cite{wasleske2023} Section 3.1, as summarized here. We employ a list of Chandra datasets, obtained by the Chandra X-ray Observatory, contained in~\dataset[DOI: 10.25574/cdc.250]{https://doi.org/10.25574/cdc.250}. 

Using the Chandra Interactive Analysis of Observations (CIAO: version 4.14), we first reprocess the observation files. By cleaning the background of the original event file, bad-pixel and a new level 2 event files are created. We then identify preliminary sources within this level 2 event file using the CIAO $\texttt{WAVDETECT}$ function. 

CIAO's $\texttt{SRCFLUX}$ software calculates new count rates and fluxes via aperture photometry. We use a 2.0" source aperture and  background aperture annulus of inner radius 20.0" and outer radius 30.0". We masked out any sources that fell within the background aperture. 

Our spectral model is $\texttt{xsphabs.abs1 * xspowerlaw.pow1}$ for all CXO data reduced. We defined the index of the photon power law $\texttt{xspowerlaw}$ as $\Gamma = 1.8$. We use the National Radio Astronomy Observatory values for galactic column density to be used in this model. This is done by setting the absorption parameter to $\texttt{abs1.nH=\%GAL\%}$. We are thus only accounting for galactic absorption, not the host galaxy absorption of each target. Deviations from a $\Gamma = 1.8$ photon power law value can happen for objects with high obscuration (see \cite{carroll2021} for example).

We measure the count and flux estimates for broad (0.5 - 7.0 keV), soft (0.5 - 1.2), medium (1.2 - 2.0 keV), hard (2.0 - 7.0 keV), and custom 2.0 - 10.0 keV and 0.5 - 8.0 keV bands. The latter custom band is to compute expected X-ray binary luminosities (\citealt{lehmer2019}; see below). We find 174 galaxies in our database having fluxes or upper limits in CXO.

\subsubsection{XMM-Newton} \label{subsubsec: Analysis - XMM}
We cross match our database with the XMM Serendipitous Source Catalog \citep{webb2020} to identify additional X-ray detections. The 4XMM-DR9 catalog is compiled from $11,204$ EPIC observations take over 19 years. \cite{watson2009} describes the pipeline from raw observation data event files from the EPIC instruments to event lists. This catalog provides nine flux bands within the range of $0.2 - 12 \text{keV}$ for its 550,000 unique sources.

Using a 2-D sky separation, we matched database objects to the closest XMM source within the XMM instrumentation resolution of 6". This lead to 45 objects with a matched source from the XMM Catalog. 17 of these objects also have a detection or upper limit from the CXO archival data.

For comparison to the \cite{lehmer2019} expected X-ray luminosity relation, we convert the 0.2 - 12 keV band into 0.5 - 8.0 keV band using the PIMMS toolkit. In the PIMMS toolkit, we assumed a power law model with index $\Gamma = 1.8$ and a Galactic absorption of $\text{n}_\text{H} = 10^{22} \text{cm}^{-2}$.

Additionally, we use the FLIX server\footnote{http://flix.irap.omp.eu/} to compute X-ray upper limits for objects that were within the fields observed by XMM but were not detected in the Source Catalog. We find 95 objects with upper limits in the XMM sky coverage.

\subsubsection{Source of the X-ray emission} \label{subsubsec: Source of X-ray Emission}

At the X-ray luminosities typically measured for dwarf galaxy nuclei, it is important to attempt to distinguish between X-ray emission from AGN, ultra-luminous X-ray sources (ULXs), and X-ray binaries (XRBs). We can estimate the expected X-ray luminosity from X-ray binaries using scaling relations. \cite{lehmer2019} used sub-galactic modeling to refine this relation for local galaxies, using a sample of galaxies with distances from $3.4$ and $29 \text{Mpc}$. Their relation follows the form 
\begin{equation}
    L_{XRB} = \alpha_{LMXB} \: \text{M}_\odot + \beta_{HMXB} \: \text{SFR}.
    \label{equation:Lehmer+2019 L_expected}
\end{equation}

This equation is the sum of two regimes of the X-ray Luminosity Function: low-mass X-ray binaries (LMXB) and high-mass X-ray binaries (HMXB). The population of LMXBs is proportional to the stellar mass of the galaxy, while the population of HMXBs is related to the SFR. 
The coefficients $\alpha_{LMXB}$ and $\beta_{HMXB}$ are empirically derived by \cite{lehmer2019}. 

We follow the procedure from \cite{birchall2020} for estimating the X-ray contribution from hot gas within the interstellar medium. \cite{mineo2012b} computes the emission from hot gas as 
\begin{equation}
    L_{Gas} = \left( 8.3 \pm 0.1 \right) \times 10^{38} \: \text{SFR} \left( \text{M}_\odot \text{yr}^{-1} \right) . 
    \label{equation:L_Gas}
\end{equation}

Both the contributions from HMXBs and hot gas rely on the galaxy SFR. To estimate the SFR, we used relations from \cite{kennicutt2012}, with corrected H$\alpha$ luminosities as our tracer. We corrected the H$\alpha$ luminosities for dust extinction using WISE band 4 ($22\mu \text{m}$) as a proxy for the $25\mu \text{m}$ correction factor. We opted to use this over near and far ultra-violet emission collected from GALEX as the SDSS fiber is smaller than the GALEX photometric resolution. We found that the UV SFR estimates were consistently greater than the corrected H$\alpha$ SFR, and that this was correlated with angular size of the galaxies.
We compute SFR estimation using
\begin{equation}
    log(\text{SFR}) = log( \text{L}_{H\alpha} + 0.020 \: \text{L}_{22 \: \mu \text{m}} ) - 41.27
    \label{equation:Kennicutt+2012 SFR}
\end{equation}
which was used in Equations \ref{equation:Lehmer+2019 L_expected} and \ref{equation:L_Gas}. 

We identify objects as having AGN in the X-ray by following the condition used in \cite{birchall2020}, 
\begin{equation}
    \frac{L_{\text{Obs.}}}{L_{XRB} + L_{Gas}} \geq 3. 
    \label{equation:x-ray AGN condition}
\end{equation}

There are 202 objects with X-ray observations in our sample; we find that 51/202 objects are X-ray AGN based on this criterion.

%%%

\subsection{Photometric Variability} \label{subsec: Analysis - Variability}

Lastly, we search for nuclear photometric variability for all galaxies with available data from the Palomar Transient Factory. In addition to the objects selected as variable in \cite{baldassare2020} based on PTF data, we construct PTF light curves for an additional 108 objects. There are 359 galaxies for which there was either no coverage in PTF, or an insufficient number of observations to construct a light curve ($<20$).

Following the technique outlined in \cite{baldassare2020}, we construct difference images using the DIAPL2 pipeline (Difference Imaging and Analysis Pipeline 2; \cite{wozniak2000}). We then carry out aperture photometry on the template and difference images with a $2.5 ''$ aperture centered on the galaxy nucleus as defined in the NASA-Sloan Atlas. The aperture size is chosen to be slightly larger than the worst seeing images used in the analysis. We use QSOFit software \citep{butler2011} to determine whether light curves are variable and whether they display AGN-like variability. AGN-like variability is determined by the goodness-of-fit to a damped random walk model \citep{kelly2009}. QSOFit outputs the parameters $\sigma_{\rm var}$, $\sigma_{\rm QSO}$, and $\sigma_{\rm not \;QSO}$ - the significance that the light curve is variable, AGN-like, or not AGN-like, respectively. Objects are selected as having AGN-like variability if $\sigma_{\rm var}$ and $\sigma_{\rm QSO}$ are greater than 2, and $\sigma_{\rm QSO}$ $>$ $\sigma_{\rm not \;QSO}$. 

We find that 8/108 new light curves have AGN-like variability in PTF, in addition to the 5 objects from \cite{baldassare2018}, 104 objects from \cite{baldassare2020}, 34 objects from \cite{burke2022} and 4 objects from \cite{wasleske2022}.

We note that we were unable to construct PTF light curves for many of the objects selected as variable AGN in \cite{burke2022}, since they are at declinations too low to be covered by PTF or are too faint for PTF's sensitivity. We still retain these objects as part of our variability sample.

 \subsection{Objects Not Selected as AGN} \label{subsec: Results - no selection}

Out of the initial 733 galaxies, 31 objects were not selected by any technique in our analysis. Six of these 31 objects were identified by coronal emission lines in \cite{salehirad2022}, which were not implemented here. 10 of these galaxies were initially identified via the the BPT or [\ion{He}{2}] diagnostics. The difference in classification for these objects may be linked to a difference in the spectrum fitting routine. Seven of these 31 objects were selected by \cite{chilingarian2018} as AGN from their broad-line emission, but eliminated by our more strict broad line criteria. 

Five of the objects were selected by \cite{sartori2015} as AGN from their mid-IR colors. There were no matches in the WISE database within 2$''$ for these objects. All five had optical spectra that lacked AGN emission line signatures.

 Finally, three objects from \cite{birchall2020} found as AGN via X-rays were not selected as AGN in our analysis. While the X-ray luminosity for these objects was taken from the XMM Source Catalog in both \cite{birchall2020} and this analysis, we use different methods to estimate SFR. Additionally, our estimate of expected XRB emission is from \cite{lehmer2019}, while \cite{birchall2020} uses the redshift-dependent version from \cite{lehmer2016}.

%%%%%%%%%%%%%%%%%%%%%%%%%%%%%%%%%%%%%%%%%%%%%%%%%%%%%%%%%%%
%%
%%      Results
%%
%%%%%%%%%%%%%%%%%%%%%%%%%%%%%%%%%%%%%%%%%%%%%%%%%%%%%%%%%%%
\section{Results} \label{sec:results}

We present the results of our analysis here. After identifying the sub-samples of objects selected by each method discussed in Section \ref{sec: Consistent Application of AGN selection tools}, we analyzed the overlap between methods. Below, we discuss the application each method to the sub-samples of AGNs found above. The percentages are representative of the fraction of galaxies at the intersection of each selection technique. Table \ref{tab: database table} summarizes our results for individual galaxies, listing each galaxy's classification according to BPT, [S II], [O I], [He II], broad lines, X-rays, IR colors, and variability diagnostics. 

Results on the intersection between methods are summarized in Figure \ref{fig:table3}.

\begin{table*}[t]
    \centering
    \textbf{AGN Selection Classification Table\\}
    \begin{tabular}{c | c c c c c c c c c c c }
\hline
Object ID& RA & Dec.& BPT & [\ion{O}{1}] & [\ion{S}{2}] & [\ion{He}{2}] & Broad-line & WISE IR & X-ray & Var.  \\
\hline
    &   (deg)   & (deg) & & & & & & & &   \\
\hline
J161756.89+225644.1 & 244.4870  & 22.9456 & AGN & Seyfert & Seyfert & AGN & -- & -- & -- & -- \\
J162539.87+404804.3 & 246.4161 & 40.8012 & AGN & Seyfert & Seyfert & AGN & AGN & -- & -- & AGN \\
J162612.35+241330.5 & 246.5515 & 24.2251 & AGN & Seyfert & Seyfert & AGN & -- & -- & -- & -- \\
J164203.17+220711.6 & 250.5132 & 22.1120 & Composite & -- & -- & -- & -- & -- & -- & -- \\
J215721.02-004342.1 & 329.3376 & -0.7284 & Composite & -- & -- & Composite & -- & -- & -- & -- \\
LEDA1133202 & 353.1052 & -0.8470 & Composite & Seyfert & Seyfert & -- & -- & -- & AGN & -- \\
LEDA2116718 & 166.4257 & 38.0565 & -- & Seyfert & Seyfert & -- & -- & -- & -- & -- \\
LEDA24384 & 130.1246 & 47.1120 & -- & -- & -- & -- & -- & AGN & -- & -- \\
LEDA2816038 & 196.7850 & 53.9623 & -- & -- & -- & -- & -- & -- & AGN & -- \\
LEDA39539 & 184.6322 & 5.8498 & AGN & LINER & Seyfert & AGN & -- & -- & AGN & -- \\

\\
\multicolumn{11}{c}{\smash{\Huge\vdots}}

\end{tabular}
    \caption{ The classification of each galaxy within our database for each AGN diagnostic. Shown here are 10 of the 733 rows. A machine readable version is included in the supplemental material. }
   \label{tab: database table}
\end{table*}
%%%

\subsection{Optical Spectroscopy} \label{subsec: Results - optical spec}

After applying optical spectroscopic selection as described in Section \ref{subsec: Analysis - emission line}, we can study the overlap between objects selected via each diagnostic and the other AGN selection methods.  

\subsubsection{BPT, [\ion{O}{1}] and [\ion{S}{2}]} \label{subsubsec: Results - BPT,OI,SII}

\begin{figure*}[t]
    \centering
    \subfigure{\includegraphics[width=\textwidth]{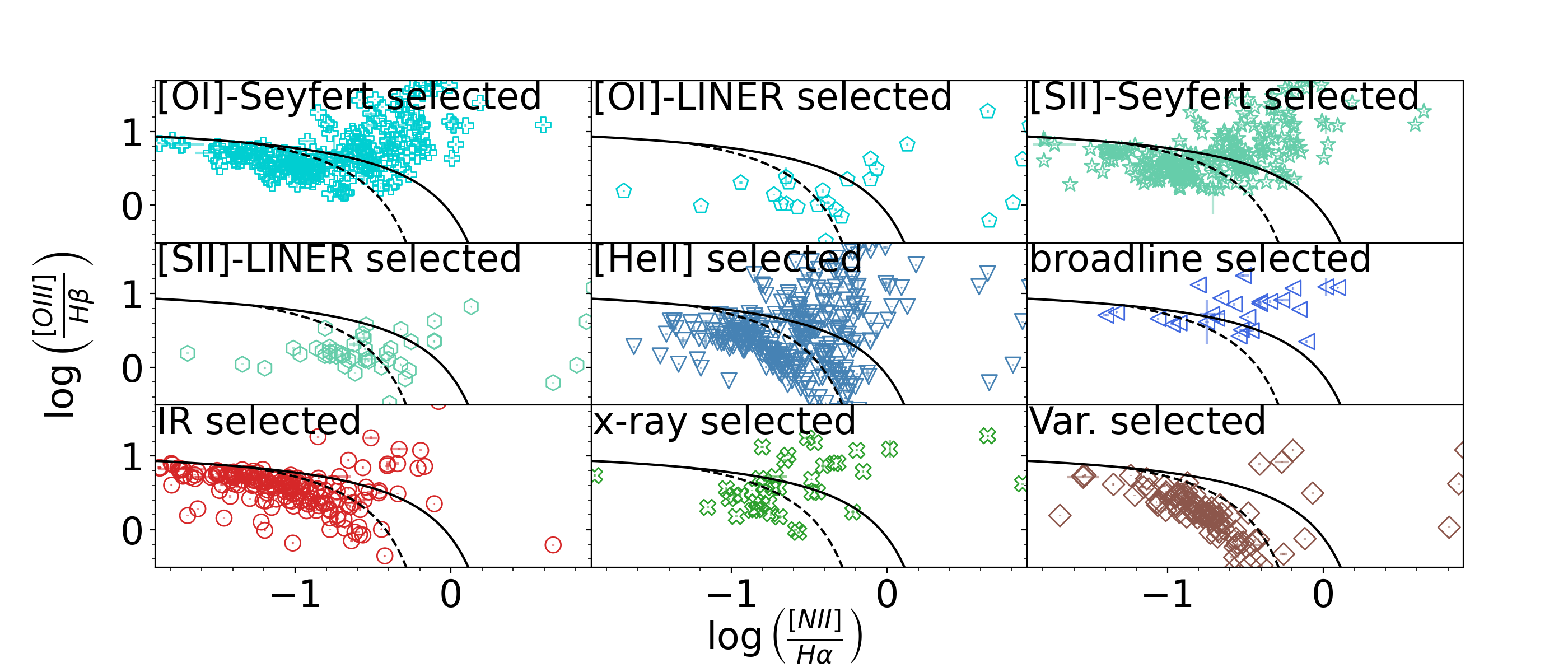}}%\quad
    \hspace{1em}% Space between image A and B
    \subfigure{\includegraphics[width=\textwidth]{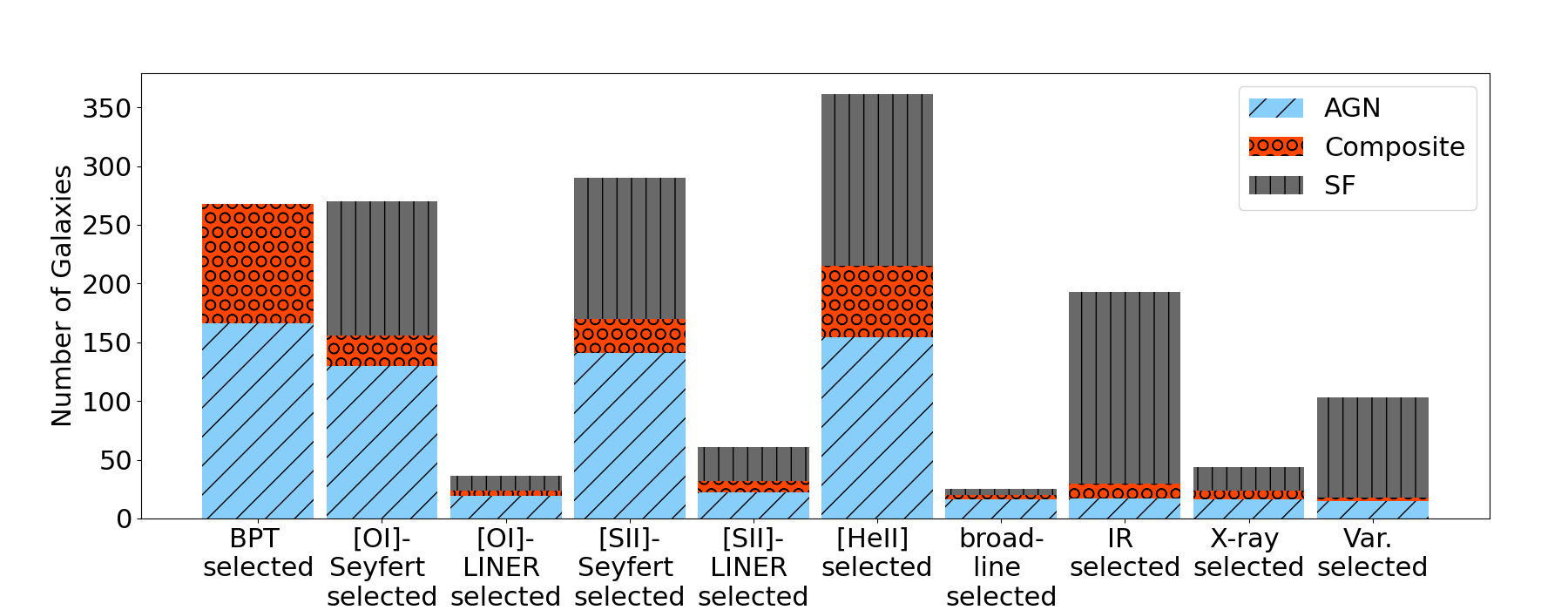}}
    \caption{ \emph{Top}: This shows the application of the BPT diagram to the AGN selected via other techniques.
    \emph{Bottom}: The results of BPT diagram applied to the subsamples of AGN selected via other techniques. We find that the BPT diagram finds evidence for BH activity in the majority of galaxies selected by other optical spectroscopic methods. Comparatively, objects identified as AGN through IR colors, X-ray emission and variability have a much lower fraction of BPT AGN and composites.
    }
    \label{fig: BPT selections and stats}
\end{figure*}

We present the application of the BPT diagnostic diagram to each of our sub-groups of AGN in top panel of Figure \ref{fig: BPT selections and stats}, with the same classifications and lines as Figure \ref{fig: BPT diagram}. We see that [\ion{O}{1}]-LINERs and [\ion{S}{2}]-LINERs have the largest scatter of their population across the BPT diagram. The IR selected population is generally located in the SF region. Some objects fall in the upper left of the SF region, possibly consistent with low-metallicity AGN (i.e. see \cite{groves2006}). Variability selected AGN are also seen mostly in the SF regime, consistent with the analysis of \cite{baldassare2020}. 

The breakdown of classifications for each of these groups of AGN is given in the bar graph of the bottom panel of Figure \ref{fig: BPT selections and stats}. The total of each bar represents the amount of  galaxies within that population that have a fitted optical spectrum. The first column gives the population of AGN found in Figure \ref{fig: BPT diagram}. We see that there is generally good agreement between the different optical spectroscopic techniques (i.e., the BPT, [\ion{O}{1}], and [\ion{S}{2}] diagrams).

\begin{figure*}[t]
    \centering
    \subfigure{\includegraphics[width=\textwidth]{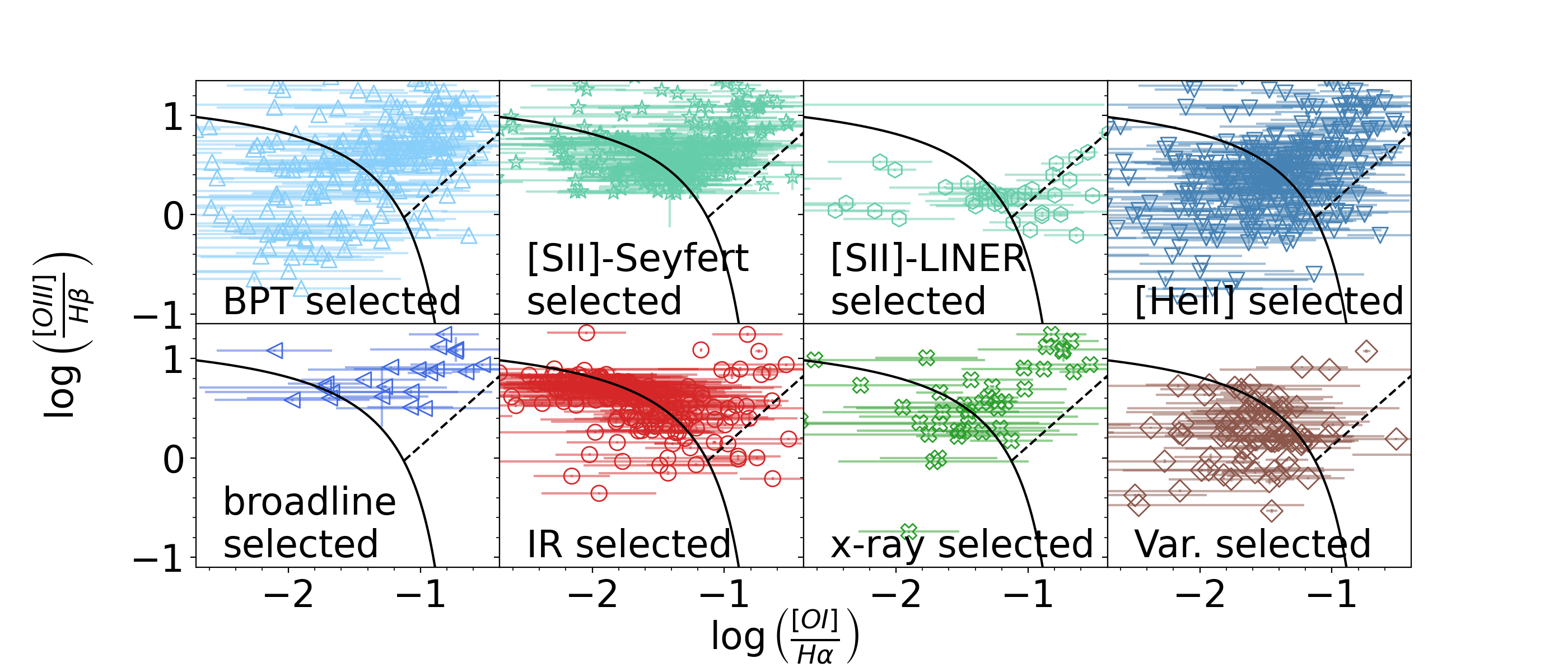}}%\quad
    \hspace{1em}% Space between image A and B
    \subfigure{\includegraphics[width=\textwidth]{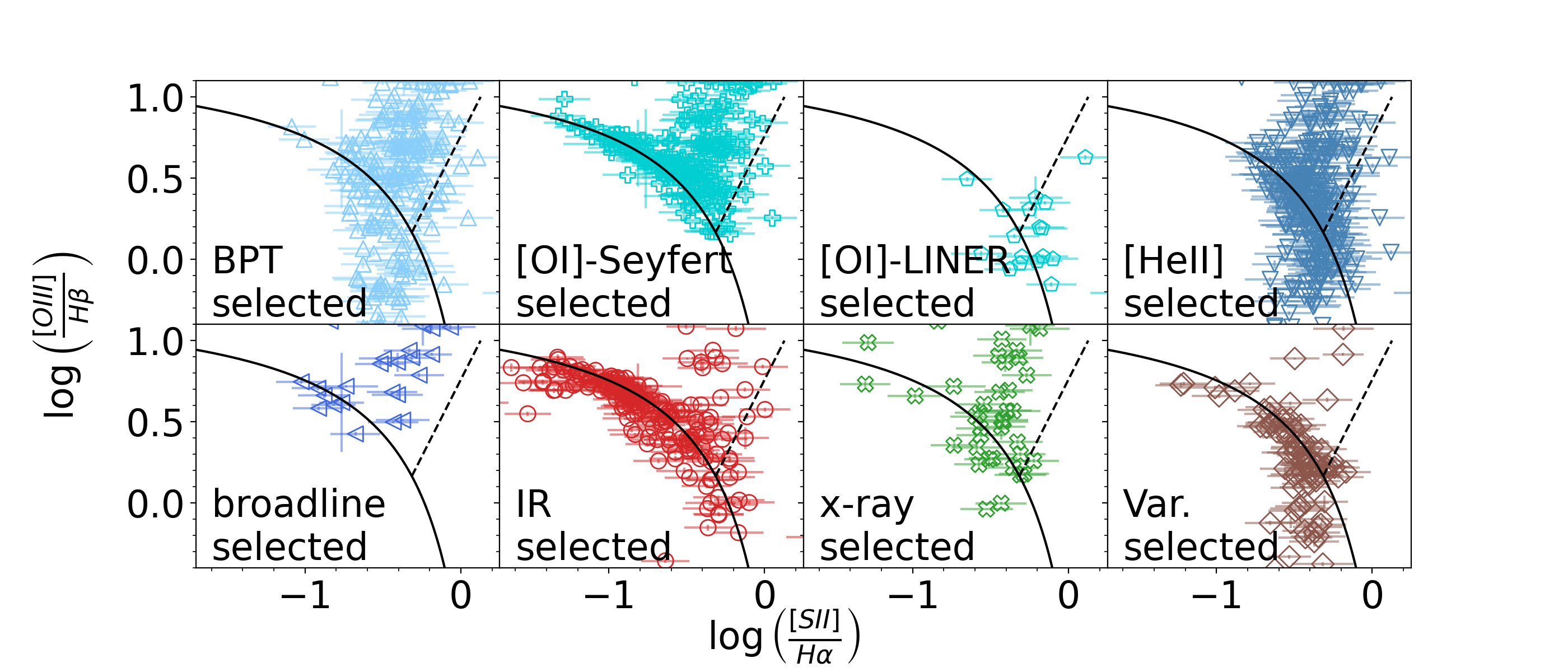}}
    \caption{ Application of the [\ion{O}{1}] (\emph{top}) and [\ion{S}{2}] (\emph{bottom}) diagrams to the populations of AGN selected by the other methods.}
    \label{fig: OI SII selections}
\end{figure*}

Furthermore, we apply the [\ion{O}{1}] and [\ion{S}{2}] diagrams to our groups of AGN, as show in Figure \ref{fig: OI SII selections}. We note that the classifications based on the [\ion{O}{1}] diagram are subject to greater uncertainties since the line flux is lower. There is also a larger scatter across the diagram compared to the [\ion{S}{2}] plot. In the top panel of the plot, we can see that the Seyferts selected by the [\ion{O}{1}] diagram overlap well with Seyferts selected by the [\ion{S}{2}] diagram, while a large fraction of BPT and [\ion{He}{2}] selected objects fall outside of the Seyfert/LINER regimes. 

In the bottom panel of Figure \ref{fig: OI SII selections}, we also find many [\ion{O}{1}]-Seyferts as [\ion{S}{2}]-Seyferts, and [\ion{O}{1}]-LINERs as LINERs. The [\ion{S}{2}] diagram overlaps better with the [\ion{O}{1}] diagram than vice versa. The majority of the IR selected population is straddling the SF line. Many variability identified objects are also not classified as AGN on the [\ion{O}{1}] or [\ion{S}{2}] diagrams.

%%%

\subsubsection{[\ion{He}{2}]} \label{subsubsec: Results - HeII}

\begin{figure*}[t]
    \centering
    \subfigure{\includegraphics[width=1.05\textwidth]{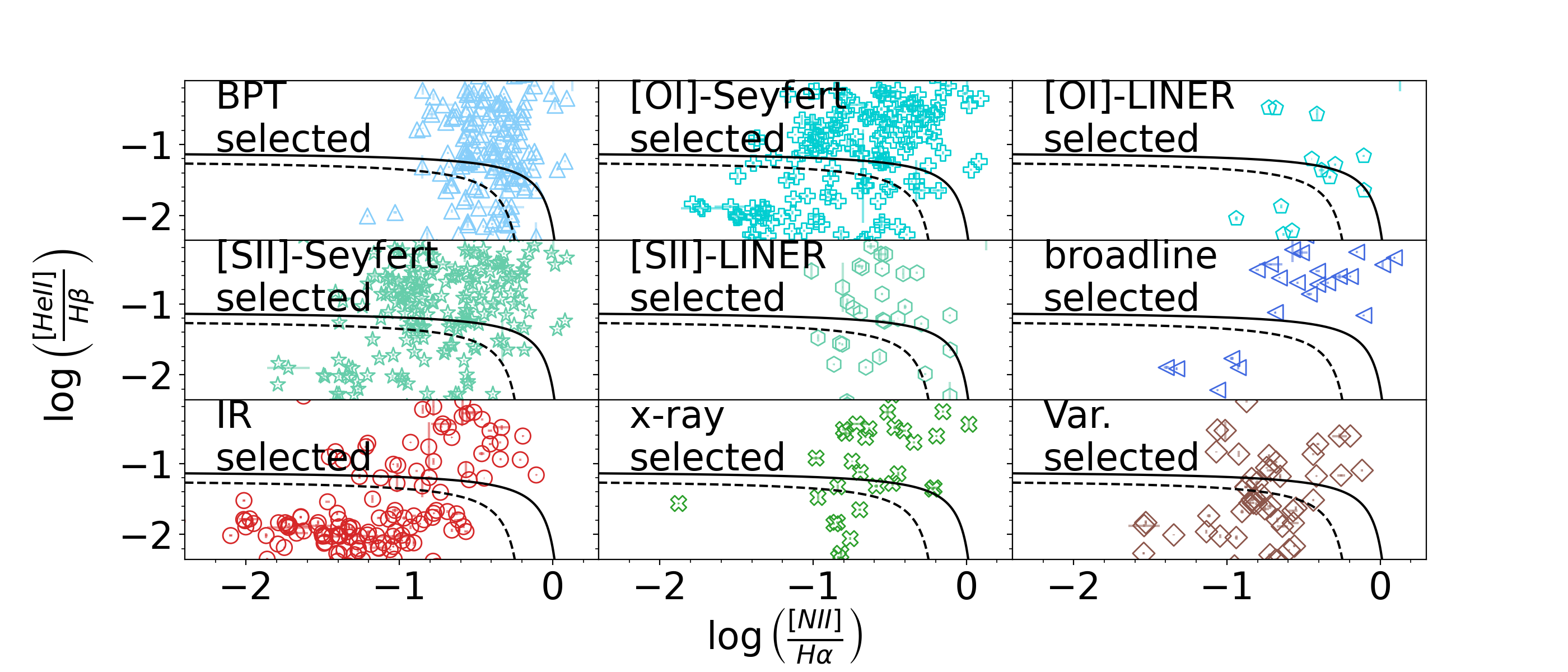}}%\quad
    \hspace{1em}% Space between image A and B
    \subfigure{\includegraphics[width=\textwidth]{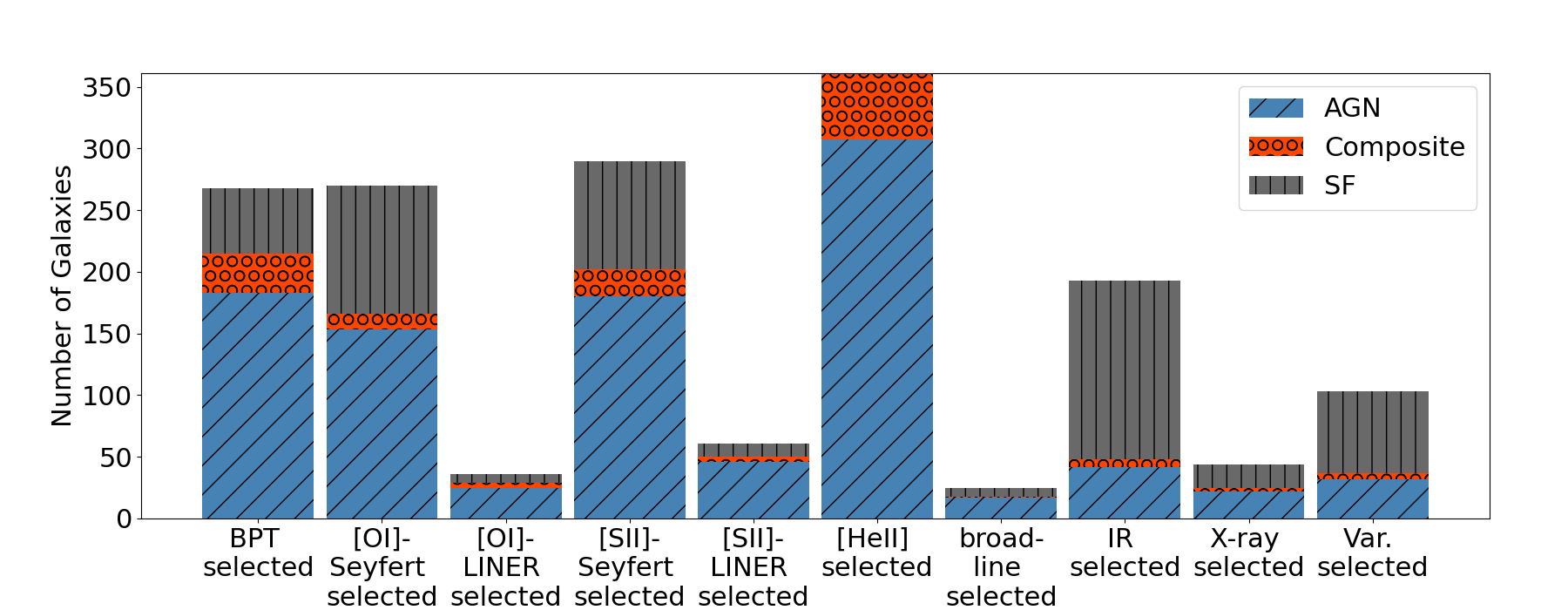}}
    \caption{ \emph{Top}: [\ion{He}{2}] emission line diagnostic applied to all sub-samples. The classifications are the same as in Figure~\ref{fig: HeII diagram}.
    \emph{Bottom}: The results of the [\ion{He}{2}] emission line diagnostic. The columns are the same as those in bottom diagram of Figure \ref{fig: BPT selections and stats}. Each bar is divided into SF, Composite, or AGN based on the region they are located in the top panel.
    }
    \label{fig: HeII selections and stats}
\end{figure*}

In Figure \ref{fig: HeII selections and stats}, we present the results of the AGN selection technique using the narrow emission line [\ion{He}{2}]. This method divides the populations into classifications of SF, Composite and AGN. In Figure \ref{fig: HeII selections and stats}, we show how the population of AGN selected by each other technique falls on the [\ion{He}{2}] diagram. In general, a small fraction of each group falls within the Composite regime of the [\ion{He}{2}] diagram. This analysis finds the majority of objects selected via the BPT, [\ion{O}{1}], and [\ion{S}{2}] diagrams are classified as [\ion{He}{2}] AGN. Similarly, the majority of X-ray and broad-line selected AGN are classified as [\ion{He}{2}] AGN. However, there is less agreement between the [\ion{He}{2}] AGN classification and variability and IR-selected AGN.

%%%
\subsubsection{Broad emission lines} \label{subsubsec: Results - broad-line}
\begin{figure*}
  \centering
    \includegraphics[width=\textwidth]{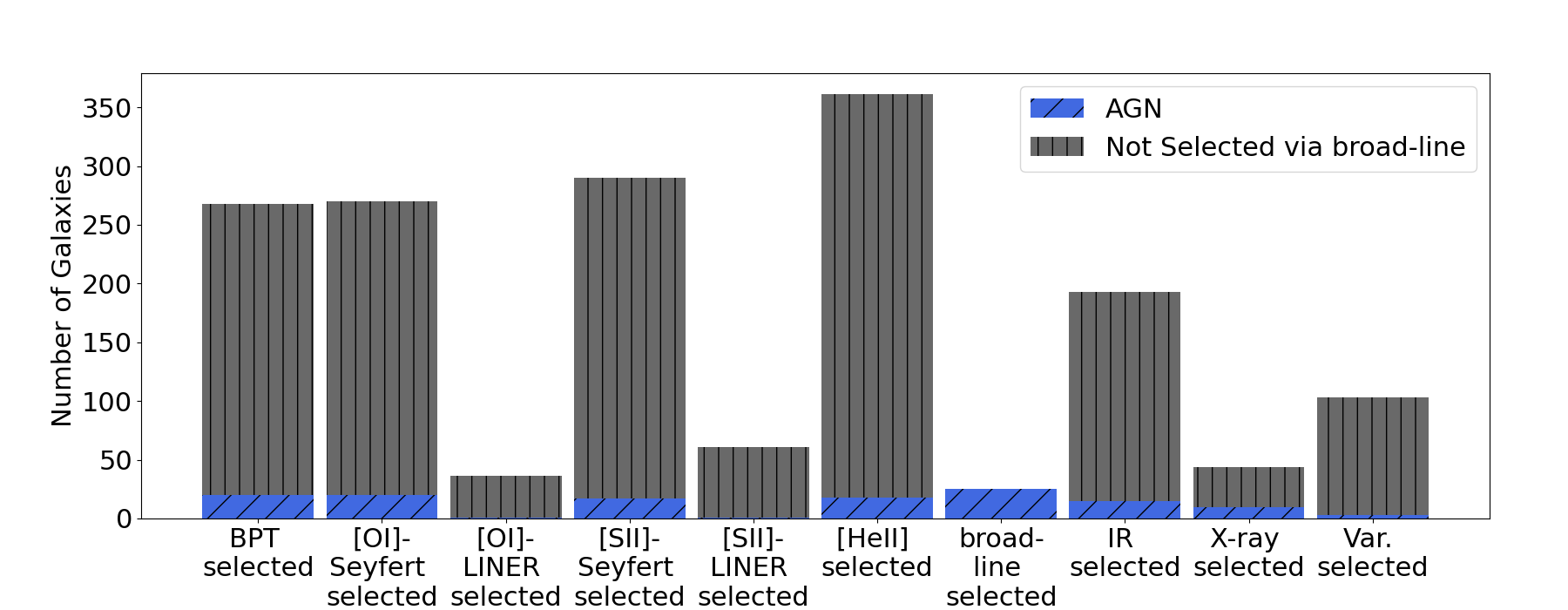}
  \caption{The cross section of the 25 broad line AGN with the other AGN selection methods. The length of each bar is the total number of fitted spectra within each group. Since only a small fraction of AGN have broad lines in general, the fraction of broad line AGN amongst the other selection techniques is small. } 
  \label{fig: broad-line stats}
\end{figure*}

Here, we study the additional AGN classifications for the 25 objects that were identified as AGN via their broad H$\alpha$ emission. In Figure \ref{fig: broad-line stats}, we show the fraction of all the other populations that comprise broad-line AGN. In general, we find that the fraction of broad-line AGN in dwarf galaxies is low. This is consistent with the fact that only a fraction of AGN in general show broad lines and the low broad H$\alpha$ luminosities for low-mass BHs. The X-ray AGN have the highest fraction of broad-line AGN ($22.7\%$), while the remaining populations have a few percent broad-line AGN. However, while the overall number of broad-line AGN is low, these objects tend to also be selected by other techniques. 

The BH masses for this sample range from $10^{4.6-7.0}\;\rm{M_{\odot}}$.
We plot the host galaxy stellar mass against the $\text{M}_{BH}$ in Figure \ref{fig: broad-line, mass to M_BH}. We include the BH mass scaling relations from \cite{reines2015} and \cite{greene2016}. The \cite{reines2015} relation is set by local ($\text{z}<0.55$) broad-line AGN. A small fraction (12/262) of the galaxies used in the Bayesian linear regression to create this relation were dwarf AGN. The relation from \cite{greene2016} is developed from high-precision BH mass measurements of water megamaser disks in spiral galaxies. Part of their population was late-type galaxies. Though there is considerable scatter, we find that the population of broad-line AGN presented here is in better agreement with the \cite{reines2015} relation. Additionally, the broad-line objects that were also selected as IR AGN reside in a relatively lower host galaxy stellar mass regime.

\begin{figure}
  \centering
    \includegraphics[width=0.52\textwidth]{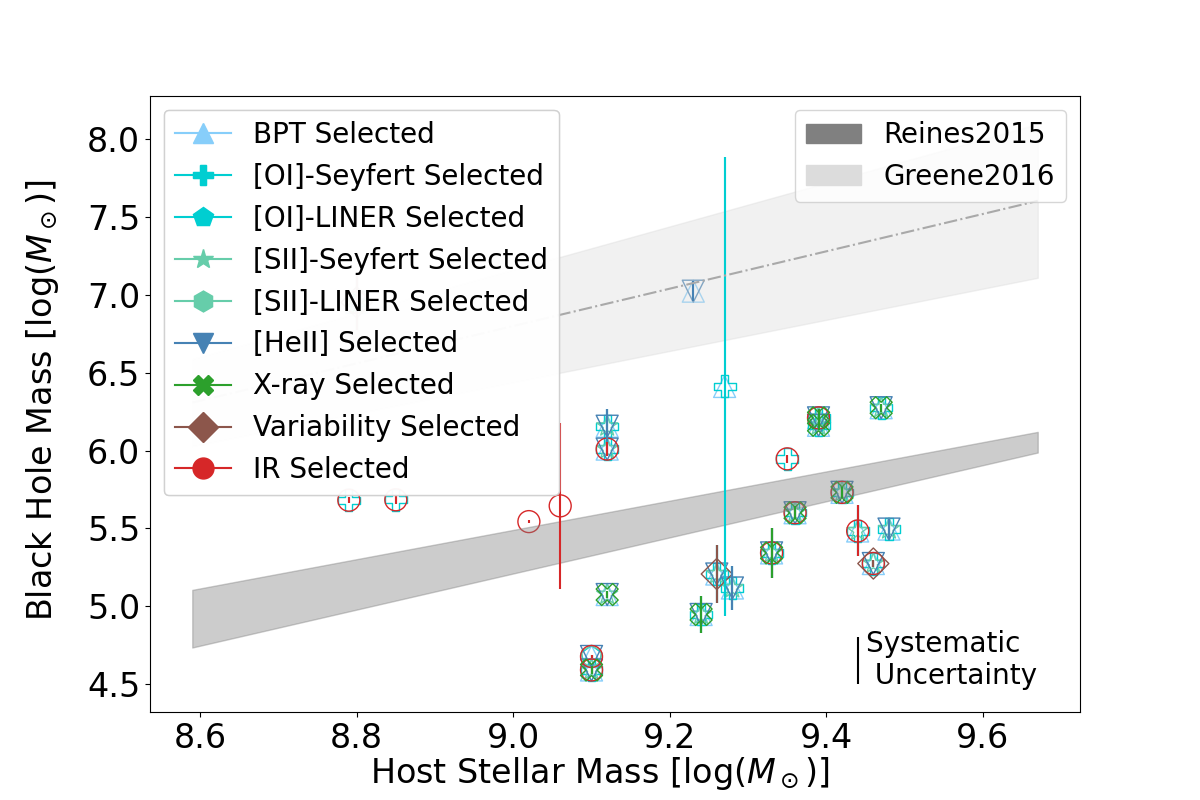}
  \caption{Black hole mass versus host galaxy stellar mass. Black hole masses are estimated using equation \ref{equation: M_BH}. Markers are set by what selection methods identified each object as an AGN, as outlined in Section \ref{sec: Consistent Application of AGN selection tools}. Error bars on the points reflect measurement uncertainties only. Systematic uncertainty of the BH mass estimate is shown by the vertical bar in the lower right. Total uncertainty in BH mass of each object is the sum of the error bar of each point plus the systematic uncertainty. The BH mass scaling relations of \cite{reines2015} and \cite{greene2016} are shown in dark and light grey, respectively. }
  \label{fig: broad-line, mass to M_BH}
\end{figure}

 %%%

\subsection{Infrared Colors} \label{subsec: Results - WISE IR}
\begin{figure*}[t]
    \centering
    \subfigure{\includegraphics[width=\textwidth]{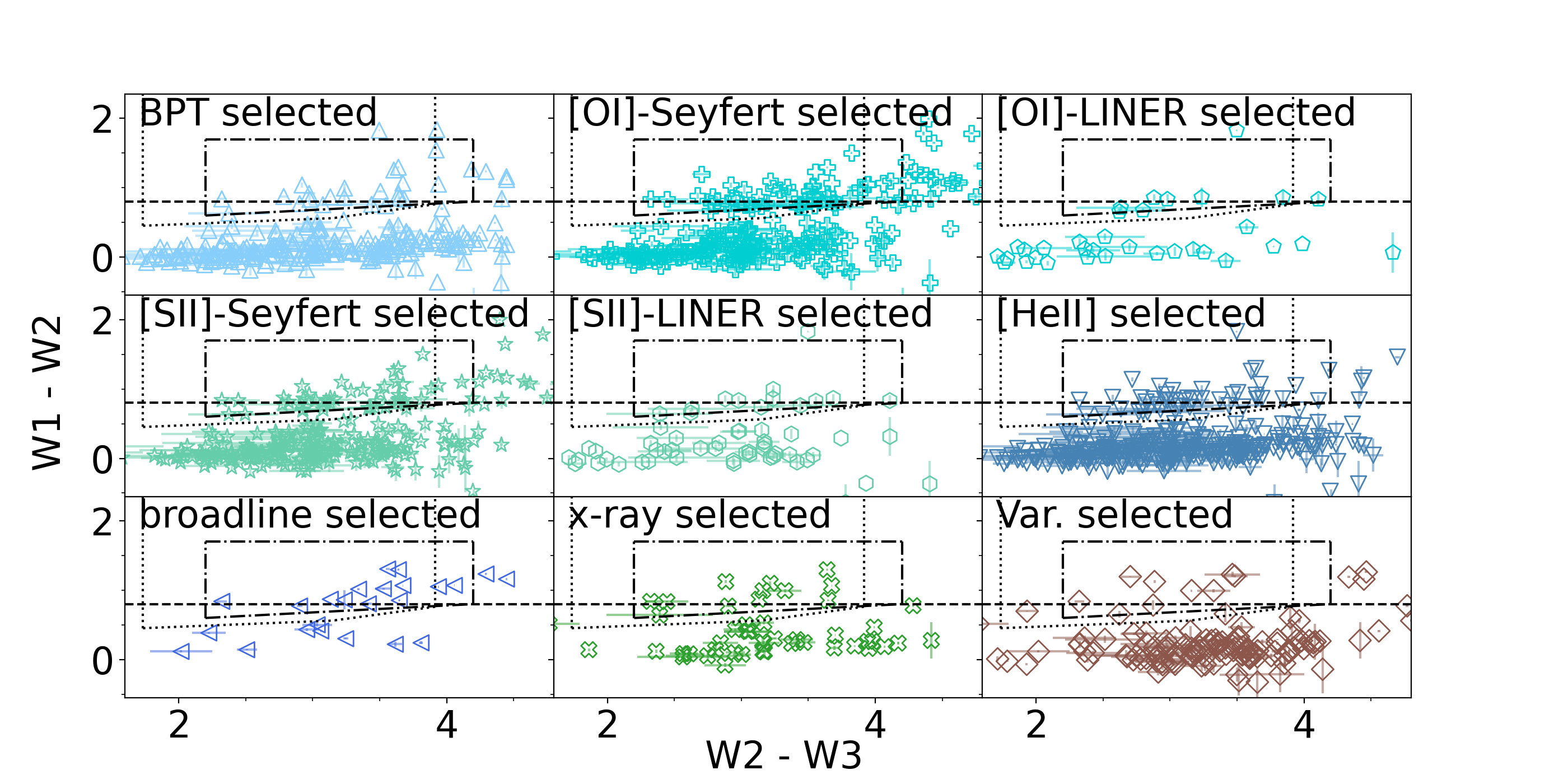}}%\quad
    \hspace{1em}% Space between image A and B
    \subfigure{\includegraphics[width=\textwidth]{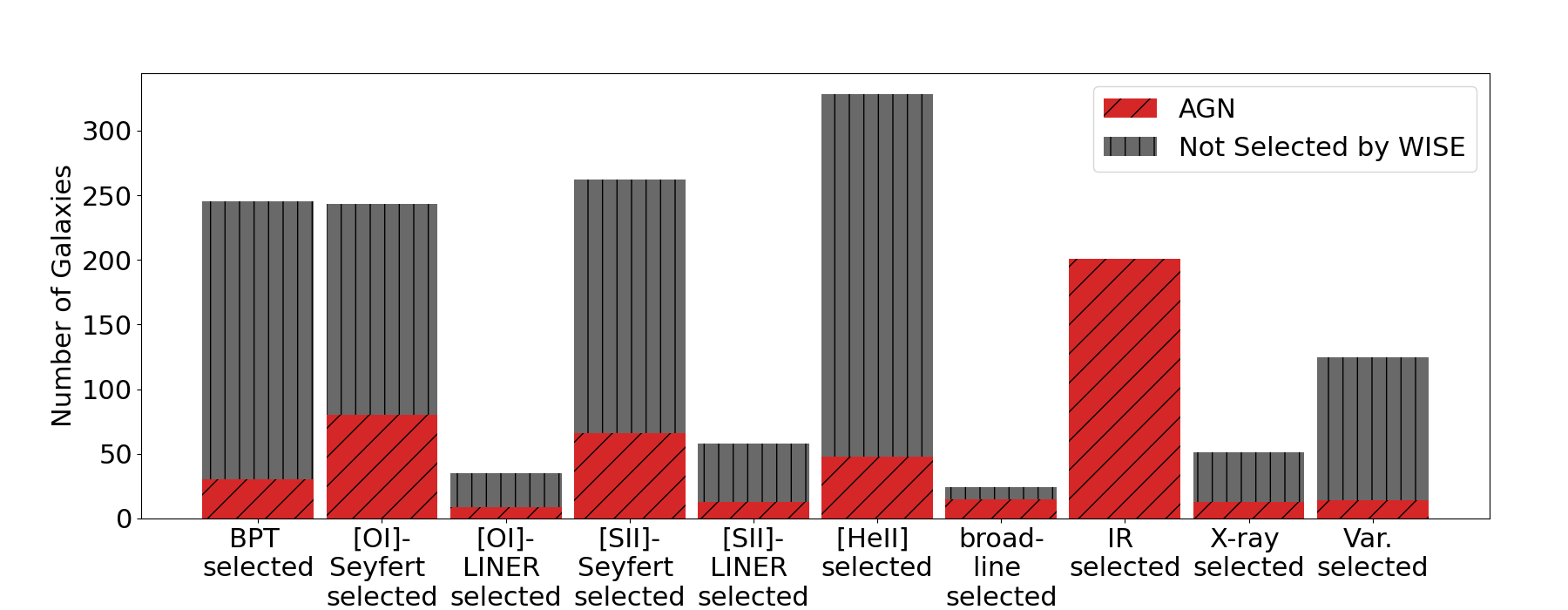}}
    \caption{ \emph{Top}: WISE color-color diagram applied to our sub-samples of AGN. 
    \emph{Bottom}: The result of the WISE color-color diagram selection. Columns are binned the same as those used in bottom diagram of Figure \ref{fig: BPT selections and stats}. We see a low overlap fraction of the IR selected sample to the other sub-samples.
    }
    \label{fig: WISE selections and stats}
\end{figure*}

Here, we study the application of the WISE color-color diagram to the other populations of AGN. We show the results in Figure \ref{fig: WISE selections and stats}. The top panel shows the color-color diagram, with the classifications defined in Section \ref{sec: Consistent Application of AGN selection tools}. The bottom plot shows a bar graph how these populations are classified by this selection method. Objects are classified as IR-AGN if they meet any of the \cite{jarrett2011, stern2012} or \cite{hviding2022} criteria. We find there is generally little overlap between IR color selection and any of the other populations, except the broad-line AGN. 15 of the 24  objects selected via their broad H$\alpha$ emission with WISE data are also selected as IR AGN. 

%%%

\subsection{X-ray emission } \label{subsec: Results - X-ray}

\begin{figure*}[t]
  \centering
    \includegraphics[width=\textwidth]{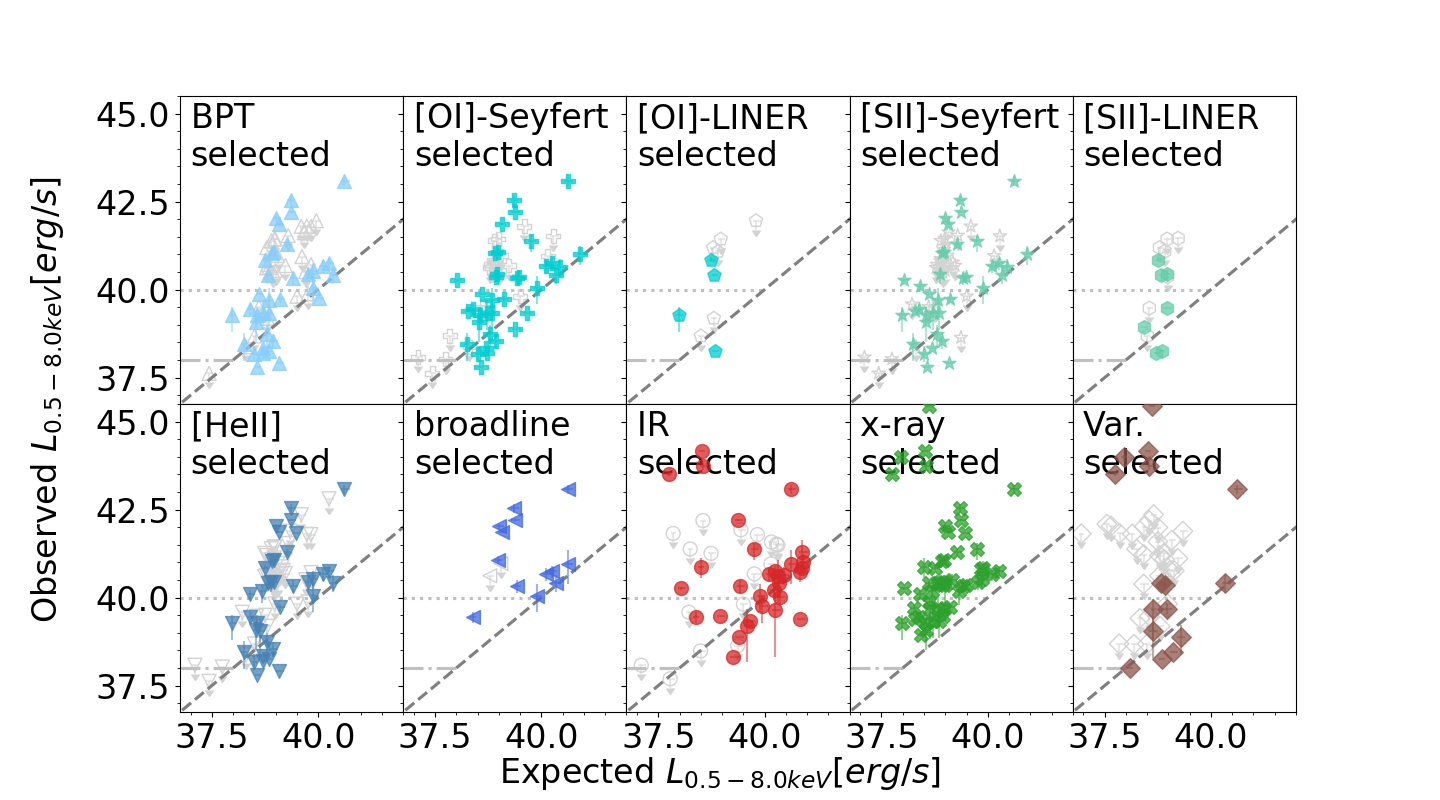}
  \caption{Expected vs. observed $0.5 - 8.0 \text{keV}$ luminosity. The expected luminosity is the sum of the estimated emission of XRB and hot interstellar medium gas from Equations \ref{equation:Lehmer+2019 L_expected} and \ref{equation:L_Gas} respectively. Points are coded based on their selection methods if they were detected in the X-ray. Upper limits are shown as light grey points in each panel. We plot a unity line to separate those objects whose observed X-ray luminosity is lower than expected, as well as detection threshold lines at $10^{38}$ and $10^{40}$ erg/s. The grey dashed line is unity. The light grey dash-dot and dotted lines are the $10^{38}$ and $10^{40}$ erg/s detection threshold lines. We also include those objects who have upper limits from their X-ray data.
  }
  \label{fig: L_exp_L_obs plot}
\end{figure*}

As described in Section \ref{subsubsec: Source of X-ray Emission} as part of our X-ray AGN selection, we estimated the expected X-ray luminosity from stars and gas within the host galaxy. We show the comparison of predicted X-ray luminosity (from XRBs and hot interstellar gas) versus the observed $0.5 - 8.0 \text{keV}$ luminosities in Figure \ref{fig: L_exp_L_obs plot}.  

We present the fractional breakdown of the X-ray classifications of each group of AGN in Table \ref{tab:xray selection table}. We distinguish between X-ray AGN (objects that have X-ray luminosities greater than three times the predicted X-ray luminosity), X-ray detections (objects that are detected, but have luminosities less than three times the predicted X-ray luminosity) and objects with X-ray upper limits. We see that optically bright and unobscured (i.e., broad-line and Seyfert) AGN have the highest fraction of X-ray AGN from this analysis. A limitation of this analysis is our inhomogeneous X-ray sample. The observations used have a wide range of sensitivities, and some of the upper limits may not be constraining.

\begin{table}[t]
    \centering
    \textbf{X-ray Sample\\}
    \begin{tabular}{c | c c c }
\hline
Selection Method & AGN    &  Detection &  Limit \\
%\hline
%    &   \%   & \% & \%  \\
\hline
\hline

\begin{tabular}[c]{@{}l@{}}BPT (AGN+Comps) \end{tabular} &  24/68   &   15/68  &  29/68 \\
\hline
\begin{tabular}[c]{@{}l@{}}{[\ion{O}{1}]} Seyferts \end{tabular} & 23/60   &   16/60   &  21/60 \\
\hline
\begin{tabular}[c]{@{}l@{}}{[\ion{O}{1}]} LINERS \end{tabular} & 3/10   &   1/10   &  6/10 \\
\hline
\begin{tabular}[c]{@{}l@{}}{[\ion{S}{2}]} Seyferts \end{tabular} & 26/67   &   14/67  &  27/67 \\
\hline
\begin{tabular}[c]{@{}l@{}}{[\ion{S}{2}]} LINERS \end{tabular} &  5/15  &   2/15   &  8/15 \\
\hline
\begin{tabular}[c]{@{}l@{}}{[\ion{He}{2}]} (AGN+Comps) \end{tabular} & 25/85 &    16/85  &      44/85  \\
\hline
\begin{tabular}[c]{@{}l@{}}Broad-line \end{tabular} & 10/15  &   3/15   &    2/15   \\
\hline
\begin{tabular}[c]{@{}l@{}}WISE IR \end{tabular} & 13/49   &   18/49  &      18/49  \\
\hline
%[c]{@{}l@{}}X-ray (AGN)\\ Total: 49 &  49/49   &  0/49    &    0/49      \\
\begin{tabular}[c]{@{}l@{}}Variability \end{tabular} &  10/69   &   13/69   &     46/69    \\

\hline

\end{tabular}
    \caption{ With a total of 202 unique objects having X-ray data across CXO and XMM surveys, we list here the fraction of objects selected as an AGN via the criteria of Equation \ref{equation:x-ray AGN condition}, or those having a detection or an upper limit to their observed broadband X-ray luminosity. Each row is divided by their selection method via our application of these methods. }
    \label{tab:xray selection table}
\end{table}

%%%

\subsection{Variability} \label{subsec: Results - variability}

\begin{figure*}
  \centering
    \includegraphics[width=\textwidth]{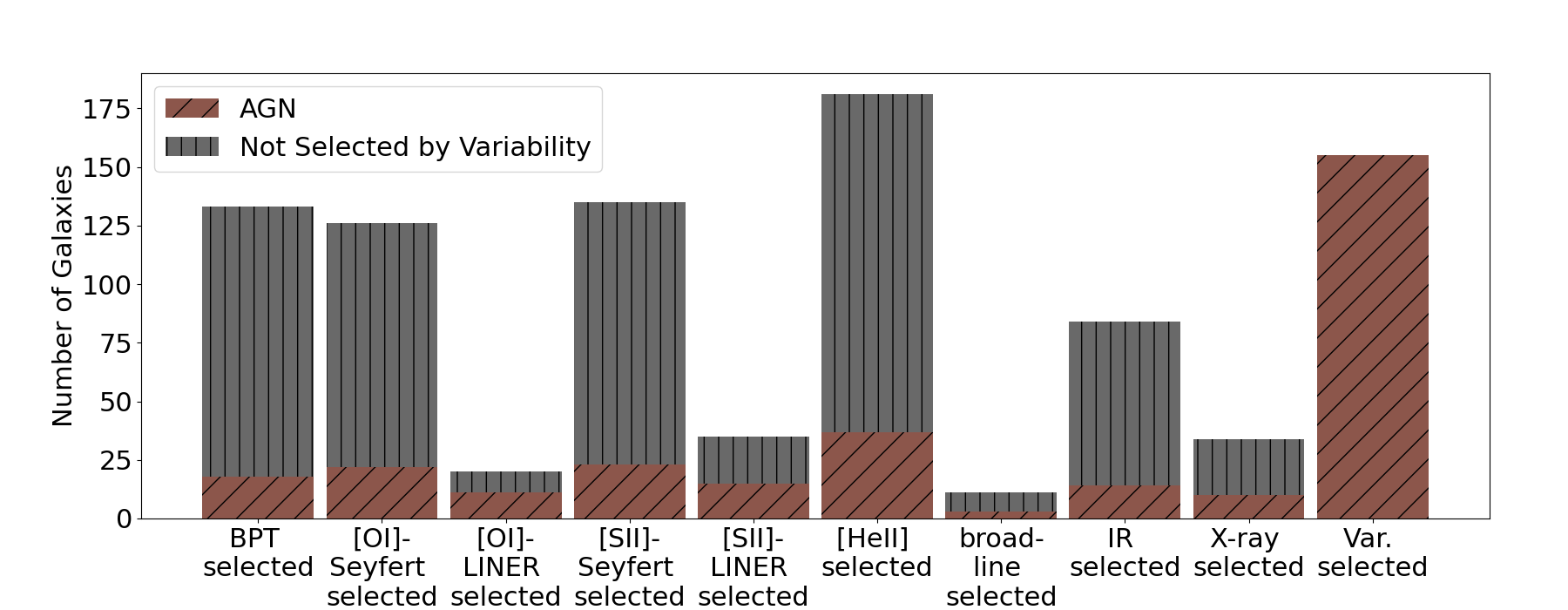}
  \caption{ The intersection of our subsets identified via optical variability. Each bar is comprised of objects for which variability data was available. In general, there is a low variability fraction amongst each of the other subsamples.
  }
  \label{fig: variability stats}
\end{figure*}

Figure~\ref{fig: variability stats} shows the fraction of each population that has AGN-like variability. Overall, variability selection seems to identify a complementary sample of AGN to most optical spectroscopic methods (last column of of the table in Figure \ref{fig:table3}). Interestingly, the [\ion{O}{1}] and [\ion{S}{2}]-selected LINERS have the highest variability fractions (55.0\% and 42.9\%, respectively). A relatively low fraction of BPT-selected AGN show variability (13.5\%), consistent with results from \cite{baldassare2018, baldassare2020}. We also see a low variability fraction amongst WISE-selected AGN ($16.7\%$). There is slightly better overlap with broad line and X-ray-selected AGN, with $25-30\%$ of objects showing AGN-like variability.

%%%%%%%%%%%%%%%%%%%%%%%%%%%%%%%%%%%%%%%%%%%%%%%%%%%%%%%%%%%
%%
%%      Discussion
%%
%%%%%%%%%%%%%%%%%%%%%%%%%%%%%%%%%%%%%%%%%%%%%%%%%%%%%%%%%%%
\section{Discussion} \label{sec:discussion}

We compiled a database of 733 AGN candidates in dwarf galaxies with stellar mass $\text{log}\left(\text{M}_* \right) \leq 9.5$. We applied a uniform set of AGN selection techniques to the entire database (when data was available), producing ten sub-samples corresponding to the objects selected as AGN by each technique. In total, 702 objects from our database were identified as AGN from our application of these selection methods. We then explored the overlap between different selection methods by measuring the percentage of objects in each sub-sample that are detected by other techniques. In the following sections, we discuss the implications of these results.

\subsection{Overlap between selection methods and implications for active fraction} \label{subsec: discussion - Selection Disagreement}

\begin{sidewaysfigure}%[htpb]
%    \centering
    \includegraphics[width=1\textwidth]{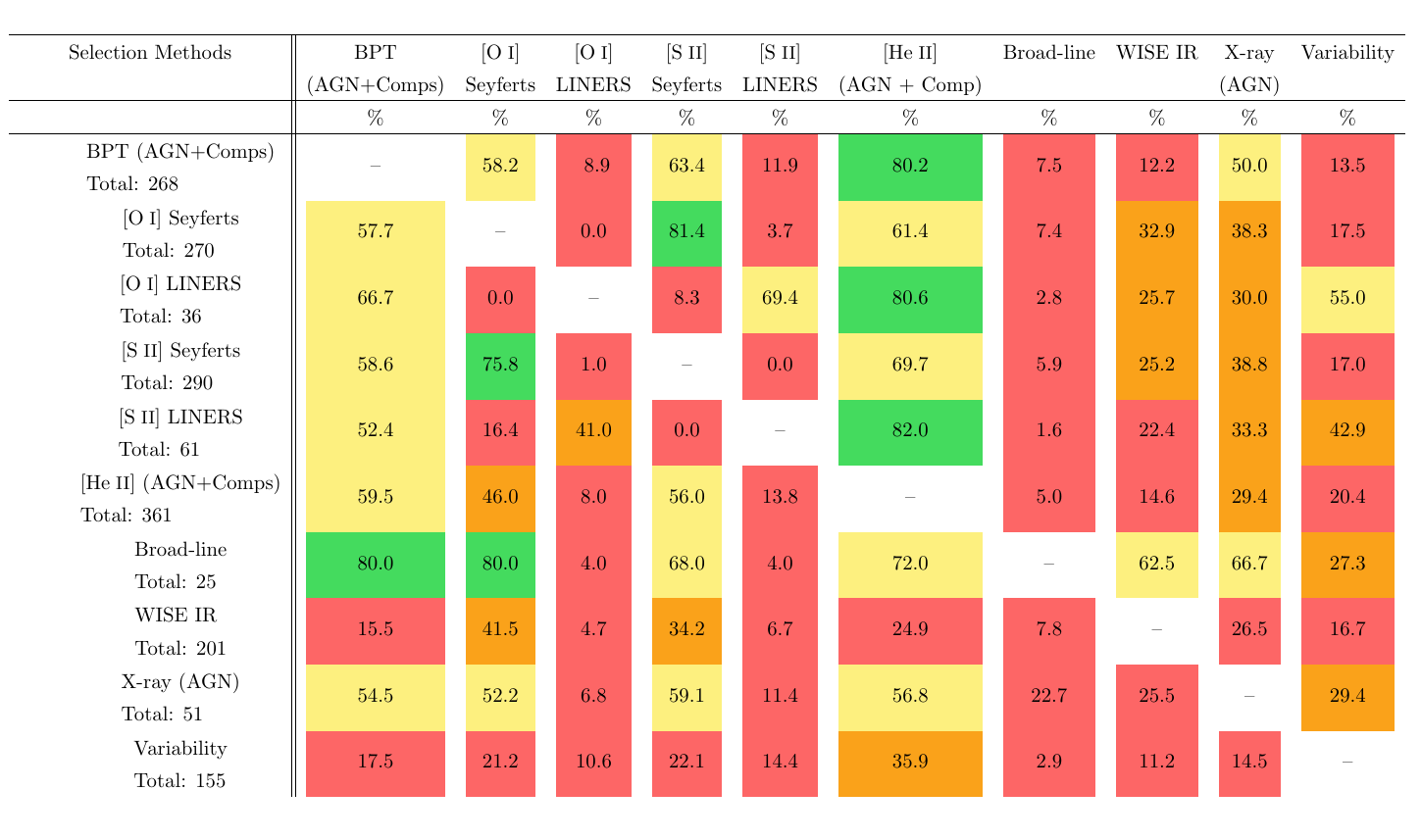}
    \caption{The number of objects from our database identified by each method is given in the first column of the table shown. Each row shows the percentage of AGN in each group that are also identified by the technique given in the column. The cells are color-coded by the percentage overlap. For an overlap in the range of 0-25\% the cell is colored red, 25-50\% in orange, 50-75\% in yellow, and 75-100\% in green.}
    \label{fig:table3}
\end{sidewaysfigure}

As shown in Figure \ref{fig:table3}, in general there is little overlap between different selection techniques. There is an overall inconsistency of the number of objects selected as AGN. The [\ion{He}{2}] diagram has the best overlap rate with other techniques with a mean of $62.6\%$, i.e., on average, [\ion{He}{2}] recovers $62.6\%$ of the AGN identified through other techniques. The method with the lowest mean overlap rate is broad-line selection at $7.1\%$. This is not unexpected, since only a fraction of all AGN show broad emission lines. It also reflects the stringency of our selection criteria and the overall low number of objects we identified as having broad lines. WISE IR and variability selection recover a low fraction of most of the AGN subsets ($\sim25\%$ for both). 

These results can be used to inform estimates of the overall AGN fraction in dwarf galaxies. Accurate estimates of the active fraction are important for constraining the overall occupation fraction of BHs, which can be used to distinguish between BH formation models. If we assume that all of the objects identified as AGN by one of the techniques are true AGN, then we can use the disagreement between techniques to estimate the missing population of AGN when using a single selection method.

For example, if a study were to only use the WISE IR colors to find their dwarf AGN candidates, they would identify only $29\%$ of the total population (201/702 = 0.29). Similarly, the BPT diagram classifies $38\%$ of objects in our sample as AGN or composite. The most complete population of AGN is identified by the [He II] diagram; 361 out of the 674 objects ($\sim53\%$) plotted on the [He II] diagram are classified as AGN.

Three quarters of the objects in our database (457/702) are selected by more than one AGN selection technique. As a caveat, not all objects have all types of data available. However, of those that are selected by only one technique, most are variability or IR color-selected. AGN with broad lines tend to be selected by most other methods. There are 14 objects that are selected by 6 or more diagnostics (out of 8 total diagnostics); 13/14 of these objects are broad-line AGN. 

Across all selection methods, we find one object that is identified by all methods as an AGN. This object, J022253.62-042929.22 \citep{burke2022} is selected as a Seyfert AGN with a estimated BH mass of $\text{log}\left(\frac{\text{M}_{BH}}{\text{M}_\odot} \right) = 6.95$. Of course, the list of objects being selected by all methods is limited to those that have optical spectroscopy, WISE data, CXO/XMM data, and variability data available. As only 25 objects met our strict criteria for a broad-line AGN, this object is one of only 2/25 that has data from all surveys we compiled. The other object, NSA 15235 is classified as an AGN in all but the X-ray category. While it is detected in X-rays (and classified as an X-ray AGN by \citealt{baldassare2017}), it did not reach our cut for X-ray classification since its measured 0.5 - 8.0 keV X-ray luminosity was $\times 1.2$ its expected luminosity. NGC 4395 (NSA 89394) is in our database and classified as an AGN in all categories for which it had data. While there was not enough PTF data to construct a light curve for NGC 4395 and assess variability, it is known to have photometric variability from previous studies \citep{burke2020}.

\subsection{Biases of AGN selection methods in dwarf galaxies}

The disagreement between methods demonstrates the need for an exploration of the relationship between the AGN population selected by each method and host galaxy properties.

Spectroscopic techniques are the most commonly applied in searching for AGN in dwarf galaxies \citep{ho1997, reines2013, moran2014, sartori2015}. However, we find that many objects identified by X-rays, variability, and IR diagnostics are not classified as AGN using optical spectroscopy. There are several possible reasons for this. 
One is dilution of the AGN by star formation. \cite{moran2002} and \cite{trump2015} both explore the dilution of AGN by star formation and find that  AGN activity can be concealed. This is particularly problematic for low-mass galaxies, where the AGN is relatively low-luminosity and the galaxy angular size is small relative to the spectroscopic fibers used by large surveys. In our sample, 600 galaxies have SDSS spectroscopy ($3''$ spectroscopic fiber) and 73 have GAMA spectroscopy (2$''$ spectroscopic fiber). These apertures contain a substantial fraction of host galaxy light. Higher-resolution optical spectroscopy around the nucleus can sometimes reveal AGN in galaxies that appear quiescent or star forming in SDSS \citep{dickey2019}.

Additionally, changes in the overall ionizing spectrum with decreasing BH mass can alter the narrow emission line ratios. \cite{cann2019} used SED modeling to demonstrate the effect of BH mass on the hardness of the spectra. As the BH mass decreases, they predicted a decrease in the ratios of [\ion{O}{3}]/H$\beta$ and [\ion{N}{2}]/H$\alpha$ resulting from an extension of the partially ionized zone.

Galaxy metallicity can also affect the emission line ratios used in optical emission line diagnostics. \cite{groves2006} show that the ratios using [\ion{N}{2}] are a robust tracer for the metallicities of these systems. In low metallicity galaxies, objects with AGN might shift out of the AGN region into the star forming region. Based on the galaxy mass-metallicity relation, this is expected to be an issue for dwarf galaxies.

IR color selection is also fraught with potential biases. Though used frequently to detect obscured AGN in higher-mass galaxies, extreme star formation in low-mass galaxies may mimic IR colors of AGN. \cite{hainline2016} showed that for galaxies in a mass range of $10^7 - 10^9\; \text{M}_\odot$, the \cite{stern2012} selection was heavily contaminated with dwarf SF galaxies. They also showed that extreme starburst can replicate the IR colors of an AGN. They conclude that while the possibility of optically SF dwarf galaxies still containing an AGN is viable, the correlation between the color-color diagram and the properties of SF demonstrates that IR emission is unlikely to be powered by AGN in this mass regime. The \cite{hviding2022} selection attempts to mitigate this issue. 

There are also challenges in distinguishing between sources of X-ray emission in dwarf galaxies, especially in low angular resolution data. Seven of the 51 objects identified as AGN via their X-ray photometry are only identified in this way. 

Recent work by \cite{thygesen2023} presents multi-wavelength high resolution data for three dwarf galaxies that were previous identified as having AGN by lower resolution X-ray imaging. They find that for two of the galaxies, the X-ray sources are off nuclear and have no companion radio emission. Their third source contains two X-ray sources separated by less than 3 arcseconds. As this source, Mrk 1434, has companion radio extended and point source, it is unresolved if this is an AGN or a ULX with an extended ``ULX bubble" radio source (i.e. see \cite{cseh2012}). This is not expected to be an issue for higher-resolution X-ray data from Chandra. 

Some ULXs may also turn out to be IMBHs and serve as laboratories to investigate the formation and evolution of BH seed mechanisms. The ULX X-ray source ESO243-49 HLX-1 is one of the best candidates for an IMBH \citep{farrell2009}. %This ULX is roughly $3.3\text{kpc}$ from the nucleus, residing in the halo of the galaxy ESO 243-49.
\cite{farrell2012} found evidence for a young stellar population around ESO243-49 HLX-1. This points to a recent merger of a tidally stripped dwarf galaxy into the main host of this system, with the ULX residing in the remnant's nuclei.

Moreover, \cite{smith2023} investigated the radio emission of ULX source CXO J133815.6+043255, residing in NGC 5252. They construct a spectral energy distribution including the new radio data to find a spectral slope that suggests emission from an unresolved radio jet. The ratio of radio-to-X-ray luminosity is shown to be consistent with radio-loud quasars and low-luminosity AGN. These results are evidence for this ULX source being a IMBH with radio jets.

With a large statistical study, \cite{tranin2023} finds that galaxies with higher star formation rates are the preferred location of ULXs. Mergers events of dwarfs into more massive hosts can spark as a results of these tidal interactions.  Thus we cannot rule out ULXs as the source of bright X-ray emission in these galaxies.
In contrast, pulsating ULX (ie. \cite{bachetti2014,furst2016,quintin2021} have been found to be pulsating neutron stars at hyper-Eddington accretion rates \citep{tranin2023}. 

AGN in dwarf galaxies may also have fundamentally different X-ray emission than in more massive galaxies. \cite{arcodia2023} finds that a sample of UV, optical, and IR-variability selected AGN in dwarf galaxies have weak X-ray emission in eROSITA observations. From  eROSITA's X-ray spectra and photometry data of ~200 objects, they detect less than 10\% of the variable AGN candidates. They rule out the significance of obscuration or variability across epochs as reasons for the lack of bright X-ray emission. They conclude that the X-ray weakness is intrinsic to the majority of their population, such as a lack of a canonical X-ray corona or a bias within variability selection. 

%%%%%%%%%%%%%%%%%%%%%%%%%%%%%%%%%%%%%%%%%%%%%%%%%%%%%%%%%%%
%%
%%      Conclusions
%%
%%%%%%%%%%%%%%%%%%%%%%%%%%%%%%%%%%%%%%%%%%%%%%%%%%%%%%%%%%%

\section{Conclusions} \label{sec:conclusions}
We compiled an initial set of 733 active dwarf galaxies found by studies using optical spectroscopic, X-ray, IR, and optical photometric variability diagnostic techniques. We applied a uniform set of AGN selection techniques to the entire sample using archival data and found 702 with AGN signatures according to our criteria. We assembled sub-samples containing galaxies identified by each selection method and cross-analyzed each of these populations with the other selection methods. Our conclusions are summarized as follows:  
\begin{itemize}

    \item Any one selection technique identifies only a fraction of the overall AGN population in dwarf galaxies. The most effective AGN selection technique is the [He II] diagram, which selected 51\% of the sample as AGN. 
    
    \item There is overall disagreement amongst AGN selection methods in the dwarf galaxy regime. The best agreement is found between methods that both use optical spectroscopy, e.g., the BPT and [He II] diagrams.

    \item The lack of  overlap between subsets of AGN emphasizes the importance of multi-wavelength observations in identifying a complete sample of active galaxies. 

    \item The smallest subset is AGN identified by broad H$\alpha$ line emission. Only $\sim4\%$ of objects in the database have broad H$\alpha$ emission. The BH masses inferred from broad H$\alpha$ range from $10^{4.6-7.0}\;\rm{M_{\odot}}$ Since only a fraction of AGN display broad lines, and the luminosity of the broad line emission scales with BH mass, this is not unexpected.  However, objects that do have broad emission lines tend to also be selected by other diagnostics. 

    \item Light curve data was available for roughly half of the database. A low fraction of BPT-selected AGN are found to be variable, consistent with previous works. Interestingly, $\sim40-50\%$ of LINERs with available light curves were found to have AGN-like variability. While variability selection is currently limited by survey depth, it is a promising method for identifying AGN missed by other techniques. Current and future variability surveys including the Vera Rubin Observatory will continue to expand objects selected via this technique.

    \item The results in this paper can be used to revise estimates of the active fraction based on only one technique for better comparison to the active fractions measured by various cosmological simulations. 
    
\end{itemize}

In a follow-up paper, we will investigate the biases of each selection technique by searching for differences in the host galaxy properties between subsets of AGN. In particular, we will examine the impacts of SFR, metallicity, obscuration, and morphology on the effectiveness of each selection technique.

%%%%%%%%%%%%%%%%%%%%%%%%%%%%%%%%%%%%%%%%%%%%%%%%%%%%%%%%%%%
%%
%%      Acknowledgments
%%
%%%%%%%%%%%%%%%%%%%%%%%%%%%%%%%%%%%%%%%%%%%%%%%%%%%%%%%%%%%
\section*{Acknowledgments} \label{sec:acknowledgments}
%\begin{acknowledgments}

This work is supported by the NASA Astrophysics Database Analysis Program through Grant 22-ADAP22-0136.  

We thank the anonymous referee for comments which have improved this manuscript. The authors are grateful to Lia Sartori for providing the relations used in her 2015 paper.

This research has made use of the SIMBAD database, operated at CDS, Strasbourg, France.

Funding for the SDSS and SDSS-II has been provided by the Alfred P. Sloan Foundation, the Participating Institutions, the National Science Foundation, the U.S. Department of Energy, the National Aeronautics and Space Administration, the Japanese Monbukagakusho, the Max Planck Society, and the Higher Education Funding Council for England. The SDSS Web Site is http://www.sdss.org/.

The SDSS is managed by the Astrophysical Research Consortium for the Participating Institutions. The Participating Institutions are the American Museum of Natural History, Astrophysical Institute Potsdam, University of Basel, University of Cambridge, Case Western Reserve University, University of Chicago, Drexel University, Fermilab, the Institute for Advanced Study, the Japan Participation Group, Johns Hopkins University, the Joint Institute for Nuclear Astrophysics, the Kavli Institute for Particle Astrophysics and Cosmology, the Korean Scientist Group, the Chinese Academy of Sciences (LAMOST), Los Alamos National Laboratory, the Max-Planck-Institute for Astronomy (MPIA), the Max-Planck-Institute for Astrophysics (MPA), New Mexico State University, Ohio State University, University of Pittsburgh, University of Portsmouth, Princeton University, the United States Naval Observatory, and the University of Washington.

This publication makes use of data products from the Wide-field Infrared Survey Explorer, which is a joint project of the University of California, Los Angeles, and the Jet Propulsion Laboratory/California Institute of Technology, funded by the National Aeronautics and Space Administration.

This research has made use of data obtained from the Chandra Data Archive and the Chandra Source Catalog, and software provided by the Chandra X-ray Center (CXC) in the application packages CIAO and Sherpa

This research has made use of data obtained from the 4XMM XMM-Newton serendipitous source catalog compiled by the 10 institutes of the XMM-Newton Survey Science Centre selected by ESA.
%\end{acknowledgments}

\software{\cite{astropy2018}}

%%%%%%%%%%%%%%%%%%%%%%%%%%%%%%%%%%%%%%%%%%%%%%%%%%%%%%%%%%%
%%
%%      Bibliography
%%
%%%%%%%%%%%%%%%%%%%%%%%%%%%%%%%%%%%%%%%%%%%%%%%%%%%%%%%%%%%
\bibliography{ewasleske}{}
\bibliographystyle{aasjournal}

%%%%%%%%%%%%%%%%%%%%%%%%%%%%%%%%%%%%%%%%%%%%%%%%%%%%%%%%%%%
%%
%%      Appendix
%%
%%%%%%%%%%%%%%%%%%%%%%%%%%%%%%%%%%%%%%%%%%%%%%%%%%%%%%%%%%%
%\appendix
%\newpage
%\pagenumbering{gobble}
%\input{table_selection_alt}
%\includepdf[pages=-,pagecommand={},width=\textwidth]{table3.pdf}

\end{document}